\documentclass{article}
\usepackage[utf8]{inputenc}
\usepackage{fullpage}
\usepackage{setspace}
\usepackage{amsmath}              
\usepackage{amssymb}
\usepackage{bm}                   
\usepackage{mathtools}            
\usepackage{color, soul}                    
\usepackage[usenames,dvipsnames]{xcolor}    
\usepackage{url, hyperref} 
\usepackage{natbib}
\usepackage{float}
\usepackage{scalerel,stackengine}  
\usepackage{array, booktabs} 
\usepackage{graphicx}
\usepackage{authblk}

\usepackage{verbatim}
\usepackage{placeins}
\usepackage{algorithm, algpseudocode}
\usepackage{comment}
\usepackage{tikz}
\usetikzlibrary{matrix,decorations.pathreplacing,calc,circuits.logic.US,circuits.logic.IEC,fit,positioning,backgrounds}
\usepackage{lscape}

\stackMath
\newcommand\reallywidehat[1]{%
\savestack{\tmpbox}{\stretchto{%
  \scaleto{%
    \scalerel*[\widthof{\ensuremath{#1}}]{\kern-.6pt\bigwedge\kern-.6pt}%
    {\rule[-\textheight/2]{1ex}{\textheight}}
  }{\textheight}%
}{0.5ex}}%
\stackon[1pt]{#1}{\tmpbox}%
}

\newcommand\independent{\protect\mathpalette{\protect\independenT}{\perp}}
\def\independenT#1#2{\mathrel{\rlap{$#1#2$}\mkern2mu{#1#2}}}

\algnewcommand{\Initialize}[1]{%
  \Statex \textbf{Initialize:}
  \Statex \hspace*{\algorithmicindent}\parbox[t]{0.95\linewidth}{\raggedright #1}
}
\algnewcommand{\Result}[1]{%
  \Statex \textbf{Result:}
  \Statex \hspace*{\algorithmicindent}\parbox[t]{0.95\linewidth}{\raggedright #1}
}

\pgfkeys{tikz/mymatrixenv/.style={decoration=brace,every left delimiter/.style={xshift=3pt},every right delimiter/.style={xshift=-1pt}}}
\pgfkeys{tikz/mymatrix/.style={matrix of math nodes,left delimiter=(,right delimiter={)},inner sep=2pt,column sep=1em,row sep=0.5em,nodes={inner sep=0pt}}}
\pgfkeys{tikz/mymatrixbrace/.style={decorate, thick}}
\newcommand\mymatrixbraceoffseth{0.8em}
\newcommand\mymatrixbraceoffsetv{0.1em}

\newcommand*\mymatrixbraceleft[4][m]{
    \draw[mymatrixbrace] ($(#1.north east)!(#1-#2-1.north east)!(#1.south east)+(\mymatrixbraceoffseth,0)$)
        -- node[right=2pt] {#4} 
        ($(#1.north east)!(#1-#3-1.south east)!(#1.south east)+(\mymatrixbraceoffseth,0)$);
}

\newcommand*\mymatrixbracebottom[4][m]{
    \draw[mymatrixbrace] ($(#1.south west)!([shift={(0.1cm,0cm)}]#1-1-#3.south east)!(#1.south east)+(0,\mymatrixbraceoffsetv)$)
        -- node[below=1pt] {#4} 
        ($(#1.south west)!([shift={(-0.1cm,0cm)}]#1-1-#2.south west)!(#1.south east)+(0,\mymatrixbraceoffsetv)$);
}

\newcommand{\textbox}[3][c]{\parbox[#1]{#2}{\strut#3\strut}}

\title{Sample size calculations for n-of-1 trials}
\author[1]{Jiabei Yang}
\author[1]{Jon A. Steingrimsson}
\author[1-2]{Christopher H. Schmid}
\affil[1]{Department of Biostatistics, School of Public Health, Brown University}
\affil[2]{Center for Evidence Synthesis in Health, School of Public Health, Brown University}
\date{\textit{\normalsize{* Correspondence: \href{mailto:jiabei_yang@brown.edu}{jiabei\_yang@brown.edu}}}}

\begin{document}

\maketitle

\textbf{Abstract:} N-of-1 trials, single participant trials in which multiple treatments are sequentially randomized over the study period, can give direct estimates of individual-specific treatment effects. Combining n-of-1 trials gives extra information for estimating the population average treatment effect compared with randomized controlled trials and increases precision for individual-specific treatment effect estimates. In this paper, we present a procedure for designing n-of-1 trials. We formally define the design components for determining the sample size of a series of n-of-1 trials, present models for analyzing these trials and use them to derive the sample size formula for estimating the population average treatment effect and the standard error of the individual-specific treatment effect estimates. We recommend first finding the possible designs that will satisfy the power requirement for estimating the population average treatment effect and then, if of interest, finalizing the design to also satisfy the standard error requirements for the individual-specific treatment effect estimates. The procedure is implemented and illustrated in the paper and through a Shiny app.
\vspace{0.3cm}


\section{Introduction}

In the presence of treatment effect heterogeneity, population average treatment effects from parallel-group randomized controlled trials may not accurately represent the risk to individual participants, making individualized treatment recommendations challenging \citep{kravitz2004evidence, greenfield2007heterogeneity, olsen2007learning}. N-of-1 trials, single participant crossover trials in which multiple treatments are given in sequentially randomized treatment periods \citep{nikles2015essential}, where we refer to a certain length of time where the same treatment is given as a treatment period, have been recommended as suitable designs for estimating individual-specific treatment effects \citep{duan2013single}.

Combining data from a series of n-of-1 trials offers several advantages compared to using data from randomized controlled trials or from a single n-of-1 trial.
Firstly, repeated measures on each participant provide more information for estimating population average treatment effects compared to measuring each participant only once in most randomized controlled trials. Secondly, because crossover trials reduce the number of participants needed to achieve the pre-specified power, a series of n-of-1 trials with multiple crossovers is even more efficient at estimating the population average treatment effect. Lastly, the ability to borrow information from other participants increases the efficiency for estimating individual-specific treatment effects compared to using only data from a single n-of-1 trial \citep{zucker2010individual, senn2019sample}. 

Designing a series of n-of-1 trials that achieve the desired power for estimating the population average treatment effect, and that, if of interest, satisfy the standard error requirements for the individual-specific treatment effect estimates requires specifying the sequences with different orders of treatments assigned to the treatment periods,
the number of participants (or trials; we can refer to trials as participants because n-of-1 trials are single participant trials and we will use ``participants'' in later development) assigned to each sequence, and the number of measurements in each period. 

Previous work on estimating sample sizes for n-of-1 trials has been limited. 
\cite{johannessen1991statistical} derived the power for estimating an individual-specific treatment effect using a single n-of-1 trial. \cite{senn2019sample} presented sample size formulae for estimating the population average treatment effect and variances of individual-specific treatment effect estimates when combining multiple n-of-1 trials. Both approach assumed specific orders of treatments assigned to treatment periods in the sequences, one measurement in each period, and independent measurements on each participant.

In this paper, we present methods for designing a series of n-of-1 trials. Section \ref{sec: design_components} defines the design components for determining the sample size of a series of n-of-1 trials. Section \ref{sec: prob_decom} presents models for estimating the population average treatment effect and the individual-specific treatment effects. Section \ref{sec: power_popavg} and \ref{sec: prec_indiv} derive the sample size formula for estimating the population average treatment effect and the standard error of the individual-specific treatment effect estimates, respectively. Section \ref{sec: find_design_components} provides details about finding the possible combinations of the design components using the derived formulae. We implement and illustrate the procedure in Section \ref{sec: simulation} and in a Shiny app described in Section \ref{sec: shiny}. Section \ref{sec: discussion} summarizes our findings and points to some future extensions.

\section{Design components for a series of n-of-1 trials} \label{sec: design_components}

\begin{figure}[htbp]
    \centering
    \includegraphics[width = 0.8\linewidth]{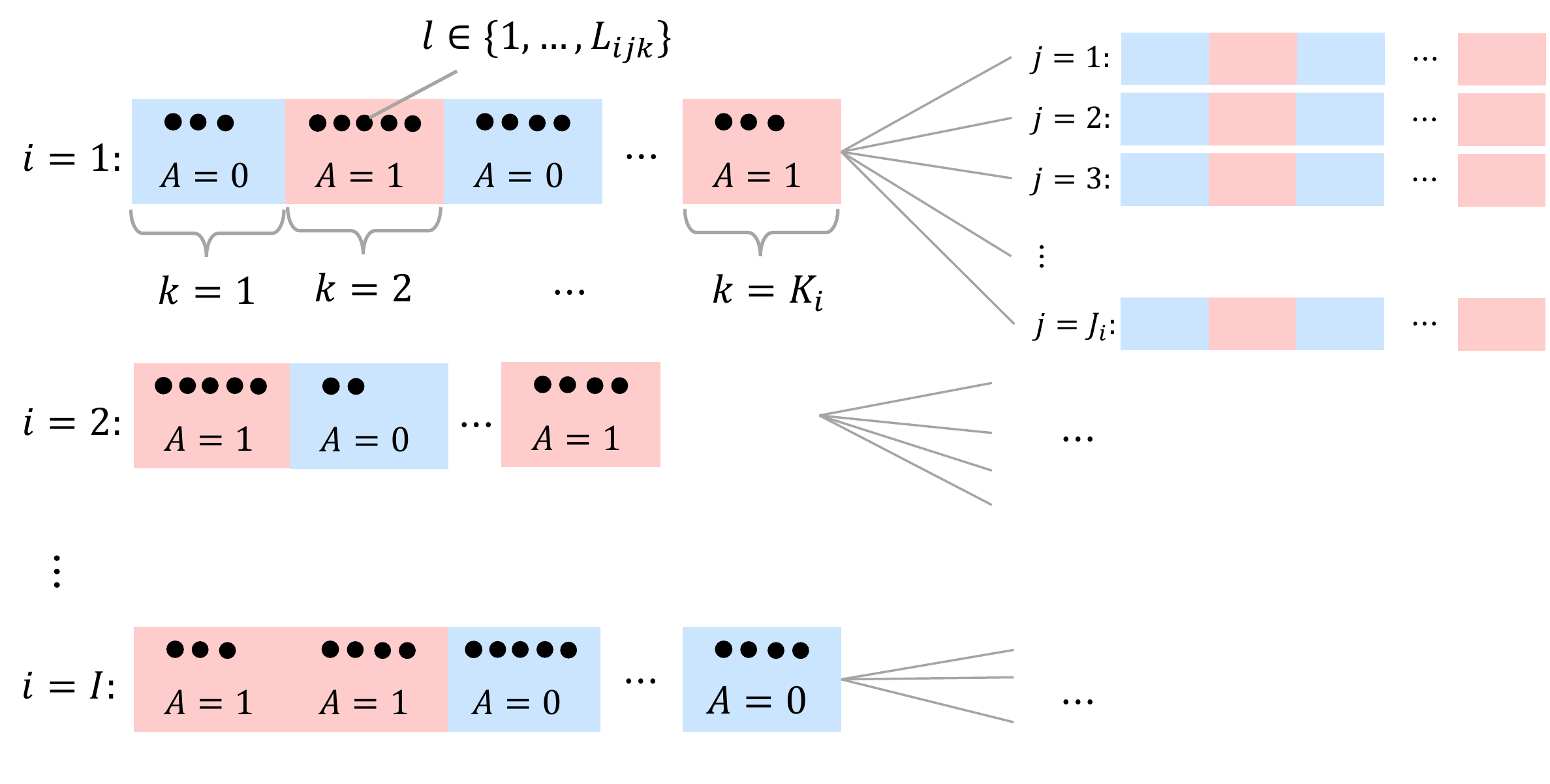}
    \caption{Indices for measurements in a series of n-of-1 trials. $A$ is the treatment assignment; we present the scenario where there are two treatments in the trials and $A$ is an indicator for being assigned to the intervention of interest. $i$, $j$, $k$  and $l$ are used to index the sequences in trials, participants assigned to each sequence, treatment periods in each sequence and measurements in each treatment period, respectively. $J_i$, $K_i$ and $L_{ijk}$ are the number of participants assigned to sequence $i$, the number of treatment periods in sequence $i$ and the number of measurements in the $k$th treatment period of participant $j$ assigned to sequence $i$, respectively.}
    \label{fig: notation}
\end{figure}

Let $Y$ be the outcome and let $A$ be the treatment assignment. Figure \ref{fig: notation} presents a schematic for the design of a series of n-of-1 trials where the outcome $Y$ is measured repeatedly at times to which treatments are assigned. 
Let $i$, $i = 1, \dots, I$, index the sequences with different orders of treatments assigned to treatment periods; let $j$, $j= 1, \dots, J_i$,
index the $J_i$ participants that are assigned to sequence $i$; let $k$, $k = 1, \dots, K_i$,
index the $K_i$ treatment periods in sequence $i$; let $l$, $l = 1,\dots, L_{ijk}$,
index the $L_{ijk}$ measurements in the $k$th treatment period for participant $j$ assigned to sequence $i$. Therefore, $Y_{ijkl}$ is the $l$th measurement on the outcome in the $k$th treatment period for participant $j$ assigned to sequence $i$ with corresponding treatment assignment $A_{ijkl}$. An n-of-1 trial design then comprises a series of trials with certain sequences where treatments are assigned to a number of treatment periods in which a number of measurements are taken. The design components consist of:
1) the sequences with different orders of treatments assigned to periods in the sequences, 
which in turn determines the number of sequences $I$ and the number of treatment periods in each sequence $K_i$, 2) the number of participants assigned to each sequence $J_i$, and 3) the number of measurements in each period $L_{ijk}$. Note that determining the sequences in the trials will determine both $I$ and $K_i$ because a different number of treatment periods leads to a different sequence. 
 
\section{Models for treatment effect estimation} \label{sec: prob_decom}

We make the following assumptions in our derivations. We assume that there are two treatments in the series of n-of-1 trials and refer to the two treatments as intervention and reference treatment; therefore, let treatment assignment $A$ be an indicator for being assigned to the intervention of interest, taking on value of 1 if assigned to the intervention and 0 if assigned to the reference treatment. We assume that the outcome is continuous. The participants are independent and the repeated measurements on each participant, indexed by $k$ and $l$, can be correlated. There is no underlying trend in the outcome or carryover effects from one treatment to the other throughout the trials. 

We now describe the models for estimating treatment effects using n-of-1 trials and start by the naive estimates for estimating the individual-specific treatment effects, which only use the data from a single participant for estimation, 
\begin{align} \label{eq: mod_indiv_naive}
    Y_{kl} = m + \delta A_{kl} + \epsilon_{kl},
\end{align}
where $m$, the intercept, is the mean of the responses when given the reference treatment, $\delta$, the slope, is the difference between the mean responses of the intervention and reference treatments, and $\epsilon_{kl}$ is the residual error. The estimate for $\delta$ gives the naive estimate for the individual-specific treatment effect. 
We specify the variance matrix for the residual error vector to account for the correlation among the repeated measurements on a single participant.

Using data from a series of n-of-1 trials, we can model the intercepts and the slopes for each participant with some structure to estimate the population average treatment effect,
\begin{align} \label{eq: mod_gen_pop}
    Y_{ijkl} = \mu_{ij} + \delta_{ij} A_{ijkl} + \epsilon_{ijkl}.
\end{align}
The intercept $\mu_{ij}$ and the slope $\delta_{ij}$ for participant $j$ assigned to sequence $i$ can then be modeled as 1) no pooling (i.e., fixed for each participant), 2) partial pooling (i.e., random from a common distribution), or 3) complete pooling (i.e., common across all participants). We usually do not fit the models with complete pooling of the intercepts because they make the strong assumption that the means of the responses on the reference treatment across the participants are the same, which usually do not hold; we also do not use the models with no pooling of the slopes for estimating the population average treatment effect because they do not give the estimate directly, but if we use the model with no pooling of both intercepts and slopes, we will have the models for each participant giving the naive estimates for the individual-specific treatment effects. Therefore, four different models span the realm of possibilities for estimating the population average treatment effect, where the intercepts can either be fixed or random and the slopes can either be random or common. The estimate for the common slope in the common slope models or for the average of the random slopes in the random-slope models estimates the population average treatment effect. 

Random slopes are more commonly used in order to allow for heterogeneous treatment effects across participants. They also suggest the shrunken estimates for the individual-specific treatment effects by borrowing information from other participants in the series of trials. Assuming a common slope implies that the individual-specific treatment effects are the same in all trials \citep{jackson2018comparison}. When the number of measurements for each trial is small, one may choose to model the population average treatment effect using a common slope for more stable estimation.

While it is common to assume random slopes, assuming random intercepts is more controversial. Some recommend fixed intercepts so that estimation of the intercepts does not bias the estimation of the population average treatment effect or of the shrunken estimates for the individual-specific treatment effects. Others have produced simulations showing that the use of fixed intercepts may bias downward the maximum likelihood estimate of the variance for the random slopes
\citep{jackson2018comparison}. Random intercepts may also be helpful in cases where the information per trial is weak and fixed intercepts are poorly estimated (e.g., cases where there are a small number of measurements per period).

We refer readers to previous studies for details on the use of random or common slopes and the use of fixed or random intercepts \citep{legha2018individual, riley2020one, white2019comparison} and will focus on the derivations using all four possible models. 

In Model \eqref{eq: mod_gen_pop}, participants are assumed to be independent, but repeated measurements on each participant may be correlated. We specify the variance matrix of the residual error vector for each participant to account for the correlation.

\section{Sample size for population average treatment effect estimation} \label{sec: power_popavg}

A general model incorporating fixed or random intercepts, random or common slopes, 
and a flexible residual error structure on each independent participant is the general linear mixed model
\begin{align} \label{eq: mod_popavg_vector}
    \bm{Y}_{ij} =& \mathbf{X}_{ij}\bm{\theta} + \mathbf{Z}_{ij} \bm{b}_{ij} + \bm{\epsilon}_{ij},\\
    \text{Var}(\bm{b}_{ij}) =& \mathbf{D}, \quad
    \text{Var}(\bm{\epsilon}_{ij}) = \bm{\Sigma}_{\bm{\epsilon}ij}, \quad \bm{b}_{ij} \independent \bm{\epsilon}_{ij}. \nonumber
\end{align}
Here, $\bm{Y}_{ij}$ and $\bm{\epsilon}_{ij}$ are the outcome and residual error vectors of length $\sum_{k=1}^{K_i} L_{ijk}$ for participant $j$ assigned to sequence $i$, $\bm{\theta}$ and $\bm{b}_{ij}$ are the fixed and random effects, and $\mathbf{X}_{ij}$ and $\mathbf{Z}_{ij}$ are the corresponding design matrices for fixed and random effects, respectively, where $i = 1, \dots, I$ and $j = 1, \dots J_i$; $\mathbf{D}$ is the variance matrix for between-individual errors and $\bm{\Sigma}_{\bm{\epsilon}ij}$ is the variance matrix 
for within-individual errors. 

Specifications of $\bm{\Sigma}_{\bm{\epsilon}ij}$ allow heterogeneous residual variance and additional correlation among the repeated measurements on each participant. The illustrations in Section \ref{sec: simulation} and the Shiny app described in Section \ref{sec: shiny} implemented commonly-used variance matrices with homogeneous residual errors and independent, exchangeable and first order autoregressive (AR-1) correlation structures. 

Table \ref{tab: mod_summ_popavg} summarizes the elements in Model \eqref{eq: mod_popavg_vector} for the four possible models that combine fixed or random intercepts with random or common slopes. We denote the fixed and random intercepts as $m_{ij}$ and $\mu + \gamma_{\mu ij}$, respectively, and the common and random slopes as $\delta$ and $\delta + \gamma_{\delta ij}$, respectively, where $\gamma_{\mu ij}$ and $\gamma_{\delta ij}$ are the random components for the random intercepts and slopes, respectively. 

The standard error for the generalized least squares estimator of the population average treatment effect \citep{laird1982random}, $\hat{\delta}$, is 
\begin{align} \label{eq: se_pop_avg}
    \text{s.e.}(\hat{\delta}) = \sqrt{ \bm{C}_{\bm{\theta}} \left( \sum_{i=1}^I \sum_{j = 1}^{J_i} \mathbf{X}_{ij}^T \mathbf{\Sigma}_{ij}^{-1} \mathbf{X}_{ij} \right)^{-1} \bm{C}_{\bm{\theta}}^T },
\end{align} 
where $\bm{C}_{\bm{\theta}}$ is the contrast matrix that pulls off the coefficient for treatment effect from the fixed effects vector or the variance for the treatment effect estimate from the corresponding variance matrix; $\mathbf{\Sigma}_{ij}$ is the variance matrix for the outcome vector $\bm{Y}_{ij}$. Table \ref{tab: mod_summ_popavg} presents the forms for $\bm{C}_{\bm{\theta}}$ and $\mathbf{\Sigma}_{ij}$ in the four possible models. 

\begin{landscape}
\begin{table}[htbp]
    \centering
    \footnotesize
    \begin{tabular}[t]{b{10mm}cccccccccccccc}
    \hline
    Model & $\mathbf{X}_{ij}$ & $\bm{\theta}$ & $\mathbf{Z}_{ij}$ & $\bm{b}_{ij}$ & $\mathbf{D}$ & $\mathbf{\Sigma}_{ij}$ & $\bm{C}_{\bm{\theta}}$\\
    \hline
    Fixed-Common & 
    \begin{tikzpicture}[mymatrixenv, baseline=0] \scriptsize 
    \matrix[mymatrix] (m)  {
        0 & \cdots & 0 & 1 & 0 & \cdots & 1 \\
        0 & \cdots & 0 & 1 & 0 & \cdots & 1 \\
        \vdots & \cdots & \vdots & \vdots & \vdots & \cdots & \vdots \\
        0 & \cdots & 0 & 1 & 0 & \cdots & 0 \\
        0 & \cdots & 0 & 1 & 0 & \cdots & 0 \\
    };
    \mymatrixbracebottom{4}{4}{Indicator column for $m_{ij}$}
    \mymatrixbraceleft{1}{2}{\textbox{1.2cm}{Treatment periods}}
    \node[text width = 1.7cm, xshift = 1 cm, yshift = 0.2cm] at (m.south east) {\scriptsize $(\sum_{k=1}^{K_i} L_{ijk}) \times (\sum_{i=1}^I J_i+1)$}; \end{tikzpicture} 
    & 
$\begin{pmatrix}m_{11}\\ 
    \vdots\\ 
    m_{IJ}\\ 
    \delta\end{pmatrix}$  
    & - & - & - & $\bm{\Sigma}_{\bm{\epsilon}ij}$ & $(\bm{0}_{\sum_{i=1}^I J_i}^T, 1)$ \\
    Fixed-Random &  
    \begin{tikzpicture}[mymatrixenv, baseline=0] \scriptsize 
    \matrix[mymatrix] (m)  {
        0 & \cdots & 0 & 1 & 0 & \cdots & 1 \\
        0 & \cdots & 0 & 1 & 0 & \cdots & 1 \\
        \vdots & \cdots & \vdots & \vdots & \vdots & \cdots & \vdots \\
        0 & \cdots & 0 & 1 & 0 & \cdots & 0 \\
        0 & \cdots & 0 & 1 & 0 & \cdots & 0 \\
    };
    \mymatrixbracebottom{4}{4}{Indicator column for $m_{ij}$}
    \mymatrixbraceleft{1}{2}{\textbox{1.2cm}{Treatment periods}}
    \node[text width = 1.7cm, xshift = 1 cm, yshift = 0.2cm] at (m.south east) {\scriptsize $(\sum_{k=1}^{K_i} L_{ijk}) \times (\sum_{i=1}^I J_i+1)$}; \end{tikzpicture}
    & 
    $\begin{pmatrix}m_{11}\\ 
    \vdots\\ 
    m_{IJ}\\ 
    \delta\end{pmatrix}$
    & \begin{tikzpicture}[mymatrixenv, baseline=0] \scriptsize
    \matrix[mymatrix] (m)  {
         1 \\
         1 \\
        \vdots \\
        0 \\
        0 \\
    };
    \mymatrixbraceleft{1}{2}{\textbox{1.2cm}{Treatment periods}}
    \node[text width = 1.1cm, xshift = 0.8cm, yshift = 0.2cm] at (m.south east) {\scriptsize $(\sum_{k=1}^{K_i} L_{ijk})$};
\end{tikzpicture} 
& $\gamma_{\delta ij}$ & $\sigma_\delta^2$ & $\mathbf{Z}_{ij} \mathbf{D} \mathbf{Z}_{ij}^T + \bm{\Sigma}_{\bm{\epsilon}ij}$ & $(\bm{0}_{\sum_{i=1}^I J_i}^T, 1)$\\
    Random-Common & 
    \begin{tikzpicture}[mymatrixenv, baseline=0] \scriptsize
    \matrix[mymatrix] (m)  {
         1 & 1 \\
         1 & 1 \\
        \vdots & \vdots \\
        1 & 0 \\
        1 & 0 \\
    };
    \mymatrixbraceleft{1}{2}{\textbox{1.2cm}{Treatment periods}}
    \node[text width = 2.1cm, xshift = 1.2cm, yshift = 0.2cm] at (m.south east) {\scriptsize $(\sum_{k=1}^{K_i} L_{ijk}) \times 2$};
\end{tikzpicture} 
& 
$\begin{pmatrix} \mu \\ \delta \end{pmatrix}$ & $\bm{1}_{\sum_{k=1}^{K_i} L_{ijk}}$ 
& 
$\gamma_{\mu ij}$ & $\sigma_\mu^2$ & $\mathbf{Z}_{ij} \mathbf{D} \mathbf{Z}_{ij}^T + \bm{\Sigma}_{\bm{\epsilon}ij}$ & (0, 1)\\
& \\
    Random-Random &
        \begin{tikzpicture}[mymatrixenv, baseline=0] \scriptsize
    \matrix[mymatrix] (m)  {
         1 & 1 \\
         1 & 1 \\
        \vdots & \vdots \\
        1 & 0 \\
        1 & 0 \\
    };
    \mymatrixbraceleft{1}{2}{\textbox{1.2cm}{Treatment periods}}
    \node[text width = 2.1cm, xshift = 1.2cm, yshift = 0.2cm] at (m.south east) {\scriptsize $(\sum_{k=1}^{K_i} L_{ijk}) \times 2$};
\end{tikzpicture}  
    & 
    $\begin{pmatrix} \mu \\ \delta \end{pmatrix}$ & $\mathbf{X}_{ij}$ 
    & 
$\begin{pmatrix}\gamma_{\mu ij} \\ \gamma_{\delta ij} \end{pmatrix}$ & $\begin{pmatrix}\sigma_\mu^2 & \sigma_{\mu, \delta}\\ \sigma_{\mu, \delta} & \sigma_\delta^2 \end{pmatrix}$ & $\mathbf{Z}_{ij} \mathbf{D} \mathbf{Z}_{ij}^T + \bm{\Sigma}_{\bm{\epsilon}ij}$ & (0, 1)\\
& \\
    \hline
    \end{tabular}
    \caption{Elements in Model \eqref{eq: mod_popavg_vector} when we use different forms of intercepts and slopes. We name the models as "form of intercept"-"form of slope"; for example, "Fixed-Common" means the model with fixed intercepts and a common slope. $\mathbf{X}_{ij}$ and $\mathbf{Z}_{ij}$ are the design matrices for fixed and random effects for participant $j$ assigned to sequence $i$ respectively; $\bm{\theta}$ and $\bm{b}_{ij}$ are the fixed effects and random effects respectively; $\mathbf{D}$ and $\bm{\Sigma}_{\bm{\epsilon}ij}$ are the variance matrices for between-individual and within-individual errors, respectively; $\bm{\Sigma}_{ij}$ is the variance matrix for the outcome vector $\bm{Y}_{ij}$; $\bm{C}_{\bm{\theta}}$ is the contrast matrix that pulls off the coefficient for treatment effect from the fixed effects vector or the variance for the treatment effect estimate from the corresponding variance matrix, respectively. The dimension of matrices or vectors are labeled to the bottom right corner.
    $\bm{0}$ and $\bm{1}$ are column vectors of $0$'s and $1$'s with the lengths labeled to the bottom right corner. $m_{ij}$ and $\mu$ are the fixed intercepts and the average of the random intercepts respectively; $\delta$ is the common slope or the average of the random slopes; $\gamma_{\mu ij}$ and $\gamma_{\delta ij}$ are the random components for the random intercepts and slopes respectively. $\sigma_\mu^2$ and $\sigma_{\delta}^2$ are the variance of the random intercepts and slopes respectively; $\sigma_{\mu, \delta}$ is the covariance between the random intercepts and slopes.}
    \label{tab: mod_summ_popavg}
\end{table}
\end{landscape}

For a given pre-specified type I error $\alpha$, 
a minimal clinically important treatment effect $\Delta$, and variance matrices for the between-individual errors when applicable, $\mathbf{D}$, and for the within-individual errors, $\mathbf{\Sigma}_{\bm{\epsilon} ij}$, 
we can achieve power of at least $1-\beta$ for estimating the population average treatment effect by finding the design components for a series of n-of-1 trials (i.e., the sequences, $J_i$, and $L_{ijk}$) that satisfy the following condition,
\begin{align} \label{eq: power}
    1-\beta \leq& \Phi \left[-Z_{1-\frac{\alpha}{2}} - \frac{\Delta}{\text{s.e.}(\hat{\delta})}\right] + \Phi \left[-Z_{1-\frac{\alpha}{2}} + \frac{\Delta}{\text{s.e.}(\hat{\delta})}\right]\nonumber\\
    =& \Phi \left[-Z_{1-\frac{\alpha}{2}} - \frac{\Delta}{\sqrt{ \bm{C}_{\bm{\theta}} \left( \sum_{i=1}^I \sum_{j = 1}^{J_i} \mathbf{X}_{ij}^T \mathbf{\Sigma}_{ij}^{-1} \mathbf{X}_{ij} \right)^{-1} \bm{C}_{\bm{\theta}}^T }}\right] \nonumber\\
    &+ \Phi \left[-Z_{1-\frac{\alpha}{2}} + \frac{\Delta}{\sqrt{ \bm{C}_{\bm{\theta}} \left( \sum_{i=1}^I \sum_{j = 1}^{J_i} \mathbf{X}_{ij}^T \mathbf{\Sigma}_{ij}^{-1} \mathbf{X}_{ij} \right)^{-1} \bm{C}_{\bm{\theta}}^T }}\right],
\end{align}
where $\Phi(\cdot)$ and $Z_{p}$ are the cumulative distribution function and the $100 \times p$th percentile of the standard normal distribution, respectively. The first term on the right hand side is small and is commonly ignored in practice.

The model we propose expands upon that of \cite{senn2019sample} in deriving sample sizes for n-of-1 trials in several ways. First, \cite{senn2019sample} assumed 
intervention and reference treatments are randomized to each pair of consecutive treatment periods and that limits the sequences to which participants can be assigned. Second, he worked with the differences between the outcomes in each pair of consecutive treatment periods where the outcome on each treatment was measured once.
Third, he assumed no correlation among the repeated measurements on each participant.  
In our formulation, we allow more flexible sequences with different orders of treatments assigned to the treatment periods, work with outcome measurements directly so the number of measurements in each period can also vary, and allow correlation among the repeated measurements on each participant.

\section{Standard error of individual-specific treatment effect estimates} \label{sec: prec_indiv}

We derive the standard error of the individual-specific treatment effect estimates in this section. Section \ref{sec: prec_naive} and \ref{sec: prec_shrunk} present the standard error of naive and shrunken estimates respectively.

\subsection{Naive estimates} \label{sec: prec_naive}

To specify the heterogeneous residual error variance and correlation among the repeated measurements on the participant of interest, we write Model \eqref{eq: mod_indiv_naive} in vector notation for participant $j^*$ assigned to sequence $i^*$ of interest, for whom we will derive the standard error of the naive estimate for the individual-specific treatment effect,
    \begin{align*}
    \bm{Y}_{i^*j^*} = \mathbf{X}_{i^*j^*}\bm{\theta}_{i^*j^*} + \bm{\epsilon}_{i^*j^*}, \quad \text{Var}(\bm{\epsilon}_{i^*j^*}) =  \bm{\Sigma}_{\bm{\epsilon}i^*j^*}.
\end{align*}
Here, $\bm{Y}_{i^*j^*}$ and $\bm{\epsilon}_{i^*j^*}$ are the outcome and the residual error vectors,
respectively,
of length $\sum_{k=1}^{K_{i^*}} L_{i^*j^*k}$,
$\mathbf{X}_{i^*j^*}$ is the design matrix with one column for intercept and the other for treatment indicators corresponding to the treatment periods 
and $\bm{\theta}_{i^*j^*} = (\mu_{i^*j^*}, \delta_{i^*j^*})^T$. Specifications of $\bm{\Sigma}_{\bm{\epsilon}i^*j^*}$ allow heterogeneous residual variance and correlation among the repeated measurements on the participant of interest.

The standard error of the resulting naive estimate for the individual-specific treatment effect, $\hat{\delta}_{i^*j^*}$, is
\begin{align} \label{eq: prec_naiv}
\text{s.e.}(\hat{\delta}_{i^*j^*}) = \sqrt{\bm{C}_{\bm{\theta}_{i^*j^*}} (\mathbf{X}_{i^*j^*}^T\mathbf{\Sigma}_{\bm{\epsilon}{i^*j^*}}^{-1}\mathbf{X}_{i^*j^*})^{-1} \bm{C}_{\bm{\theta}_{i^*j^*}}^T},
\end{align}
where $\bm{C}_{\bm{\theta}_{i^*j^*}} = (0, 1)$.

The naive estimates have limitations which show the advantage of the shrunken estimates. At least two treatments are required to estimate the treatment effect and standard error using the data from a single participant. 
Additionally, estimating $\bm{\Sigma}_{\bm{\epsilon}_{i^*j^*}}$ using data from only one participant results in considerable uncertainty \citep{senn2019sample}.

\subsection{Shrunken estimates} \label{sec: prec_shrunk}

Model \eqref{eq: mod_gen_pop} suggests the shrunken estimate for the individual-specific treatment effect for a given participant $j^*$ assigned to sequence $i^*$, $\hat{\delta}_{i^*j^*} = \hat{\delta}+\hat{\gamma}_{\delta i^*j^*}$.
Appendix \ref{app: prec_shrunk} shows that the variance of $\hat{\delta}_{i^*j^*} - \delta_{i^*j^*}$, which accounts for the additional variation in random treatment effects \citep{laird1982random}, is
\begin{align} \label{eq: prec_shrunk}
    &\text{Var}(\hat{\delta}_{i^*j^*} - \delta_{i^*j^*}) \nonumber\\
    =& \bm{C}_{\bm{\theta}} \left( \sum_{i = 1}^I \sum_{j = 1}^{J_i} \mathbf{X}_{ij}^T \mathbf{\Sigma}_{ij}^{-1} \mathbf{X}_{ij} \right)^{-1} \bm{C}_{\bm{\theta}}^T -2\bm{C}_{\bm{\theta}} \left[ \sum_{i = 1}^I \sum_{j = 1}^{J_i} \mathbf{X}_{ij}^T \mathbf{\Sigma}_{ij}^{-1} \mathbf{X}_{ij} \right]^{-1} \mathbf{X}_{i^*j^*}^T \mathbf{\Sigma}_{i^*j^*}^{-1} \mathbf{Z}_{i^*j^*} \mathbf{D}  \bm{C}_{\bm{b}}^T + \nonumber\\
    & \bm{C}_{\bm{b}} \left[\mathbf{D} - \mathbf{D} \mathbf{Z}_{i^*j^*}^T \mathbf{\Sigma}_{i^*j^*}^{-1} \mathbf{Z}_{i^*j^*} \mathbf{D} + \mathbf{D} \mathbf{Z}_{i^*j^*}^T \mathbf{\Sigma}_{i^*j^*}^{-1} \mathbf{X}_{i^*j^*}\left( \sum_{i=1}^I \sum_{j = 1}^{J_i} \mathbf{X}_{ij}^T \mathbf{\Sigma}_{ij}^{-1} \mathbf{X}_{ij} \right)^{-1} \mathbf{X}_{i^*j^*}^T \mathbf{\Sigma}_{i^*j^*}^{-1} \mathbf{Z}_{i^*j^*} \mathbf{D}\right]\bm{C}_{\bm{b}}^T
\end{align}
where $\bm{C}_{\bm{b}}$ is $1$ for the model with fixed intercepts (i.e., Fixed-Random in Table \ref{tab: mod_summ_popavg}) and is $(0, 1)$ for the model with random intercepts (i.e., Random-Random in Table \ref{tab: mod_summ_popavg}); the remaining terms are as defined in Table \ref{tab: mod_summ_popavg}. The square root of the variance gives the standard error for the shrunken estimate.

\section{Finding Design components in a series of n-of-1 trials} \label{sec: find_design_components}

We recommend that practitioners take the following two steps when designing n-of-1 trials. 
Firstly, use the sample size formula for estimating the population average treatment effect to find the possible combinations of the design components that achieve the pre-specified power requirement; secondly, if they are further interested in estimating the individual-specific treatment effects, they will finalize the design by picking the combinations of the design components that also satisfy the standard error requirements for the individual-specific treatment effect estimates. 
Illustrations in Section \ref{sec: simulation} and the Shiny app described in Section \ref{sec: shiny} will follow these two steps.
In this section, we discuss the ways to find the possible combinations of the design components using the formulae in Section \ref{sec: power_popavg} and \ref{sec: prec_indiv}.

Among the design components, the sequences with different orders of treatments assigned to periods in the sequences, which in turn determines the number of sequences $I$ and the number of treatment periods in each sequence $K_i$, the number of participants assigned to each sequence $J_i$, and the number of measurements in each treatment period $L_{ijk}$, where $i = 1, \dots, I$, $j = 1, \dots, J_i$ and $k = 1, \dots, K_i$, the first step is to determine the sequences in the series of trials. 

Sequences can be specified in two general ways, manually or by randomization. Manual determination usually pre-specifies sequences for a specific purpose.
For example, one might wish to start with a placebo, followed by an intervention and then alternate these two treatments a certain number of times. Or one might 
constrain the number of times the same treatment could be given consecutively when choosing sequences. 

We can also follow randomization schemes to determine the sequences.
Randomization schemes include but are not limited to: 1) alternating sequences, where the two treatments alternate with the first period assigned at random, 2) pairwise randomization, randomly allocating the order of the two treatments in each consecutive pair of treatment periods, where we refer to the consecutive pairs of treatment periods as blocks, 3) restricted randomization, randomly assigning treatments in the sequence with the restriction that 
each treatment is assigned to the same number of periods,
or 4) unrestricted randomization, completely randomizing treatments to treatment periods in the sequence \citep{johannessen1991statistical}. 

If the number of treatment periods in the sequences are the same (i.e., $K_1 = \cdots = K_I = K$), we can derive the number of sequences $I$ from the number of treatment periods $K$ in each sequence following the randomization schemes (Table \ref{tab: numseq_k}). Except under unrestricted randomization,
each treatment is generally assigned to the same number of treatment periods in each sequence so all the sequences in a two treatment design will have an even number of treatment periods. 
Here, we include odd number of treatment periods for completeness. For odd number of treatment periods in each sequence: 1) under pairwise randomization, we randomly allocate the order of the two treatments in each consecutive pair of periods in the first $K-1$ periods and randomly assign a treatment to the last period; 2) under restricted randomization, we randomly assign treatments with the restriction that there is only one period difference between the number of periods assigned with the two treatments, regardless of the direction.

\begin{table}[htbp]
    \centering
    \begin{tabular}{lcccccccc}
        \hline
         Randomization scheme & Odd $K$ & Even $K$ \\
         \hline
         Alternating sequences & \multicolumn{2}{c}{2}\\
         Pairwise randomization & $2^{\frac{K+1}{2}}$ & $2^{\frac{K}{2}}$\\
         Restricted randomization & $2\binom{K}{(K-1)/2}$ & $\binom{K}{K/2}$\\
         Unrestricted randomization & \multicolumn{2}{c}{ $2^{K}$ }\\
         \hline
    \end{tabular}
    \caption{The number of sequences $I$ under different randomization schemes given the number of treatment periods $K$ in the sequence. "Odd $K$" and "Even $K$" refer to scenarios where there are odd and even number of treatment periods in the sequences respectively. When there are odd number of periods in each sequence: 1) under pairwise randomization, we randomly allocate the order of the two treatments in each consecutive pair of crossover periods in the first $K-1$ periods and randomly assign a treatment to the last period; 2) under restricted randomization, we randomly assign treatments with the restriction that there is only one period difference between the number of periods assigned with the two treatments, regardless of the direction.}
    \label{tab: numseq_k}
\end{table}

We will be able to determine the number of sequences $I$ and the number of treatment periods $K_i$ when sequences are specified manually and the relationship between $I$ and $K$ when following randomization schemes for designs with the same number of periods across sequences. The next step is to determine the number of participants assigned to each sequence $J_i$ and the number of measurements in each treatment period $L_{ijk}$. 

We focus on balanced designs in which the number of participants assigned to each sequence, $J_i$, the number of treatment periods in each sequence, $K_i$, and the number of measurements in each treatment period, $L_{ijk}$, are the same so that $J_i=J$, $K_i=K$, and $L_{ijk}=L$. Unbalanced designs where there can be different numbers of treatment periods in the sequences, different numbers of participants assigned to the sequences, or different numbers of measurements in the treatment periods are much more complex and have more moving parts making optimization difficult. Furthermore, if balance is not desired, the nature of the imbalance is often specified. It will be possible to fix one or more of the design components and solve a constrained optimization problem.

Because both the number of participants, $IJ$ and the number of repeated measurements on each participant, $KL$, affect the cost and practicality of the series of trials \citep{senn2002cross}, determining $J$ and $L$ once $I$ and $K$ are determined trades off between the number of participants and the number of repeated measurements per participant. We can therefore fix either $KL$ or $IJ$ and then find the other.

Specifically, if we choose to first fix $KL$, $I$ and $K$ are determined when investigators specify sequences manually, and $L$ will be fixed because $K$ is determined; when following randomization schemes we find the possible combinations of $K$ and $L$ that lead to the fixed product and $I$ will be determined by $K$ following the relationship in Table \ref{tab: numseq_k}. We will then be able to calculate the possible values for the number of participants assigned to each sequence, $J$, using Equation \eqref{eq: power} in both scenarios. Alternatively, if we choose to first fix $IJ$, following a similar procedure, we will also be able to calculate the possible values for the last element in this case, the number of measurements in each treatment period, $L$. These calculations find the possible combinations of the design components that satisfy the power requirement for estimating the population average treatment effect. 

One can then finalize the design using Equations \eqref{eq: prec_naiv} or \eqref{eq: prec_shrunk} to calculate the standard errors of the individual-specific treatment effect estimates if these are of interest and choose the designs that also achieves the required precision.

\section{Comparison of designs} \label{sec: simulation}

To illustrate the trade-offs between the number of participants and the number of measurements per participant, we consider a balanced design with pairwise randomization, probably the most common type of n-of-1 design. In balanced designs, the dimensions of within-individual error variance matrices, $\bm{\Sigma}_{\bm{\epsilon}ij}$'s, are the same across participants. We assume that the variance matrices do not vary by participants and equal to $\bm{\Sigma}_{\bm{\epsilon}}$. Additionally, 
the residual errors are homogeneous with a variance of 4 and have an AR-1 correlation structure with a correlation coefficient of 0.4. We set a minimal clinically important treatment effect of $\Delta = 1$, Type I error rate of $\alpha = 0.05$ and Type II error rate of $\beta = 0.2$. When applicable in the model, the variance of the random intercepts ($\sigma_{\mu}^2$) and slopes ($\sigma_{\delta}^2$) are 4 and 1 respectively, and the covariance of the random intercepts and the random slopes ($\sigma_{\mu, \delta}$) is 1 (i.e., the correlation between the random intercepts and random slopes is 0.5). 

\begin{figure}[htbp]
    \centering
    \includegraphics[width=\textwidth]{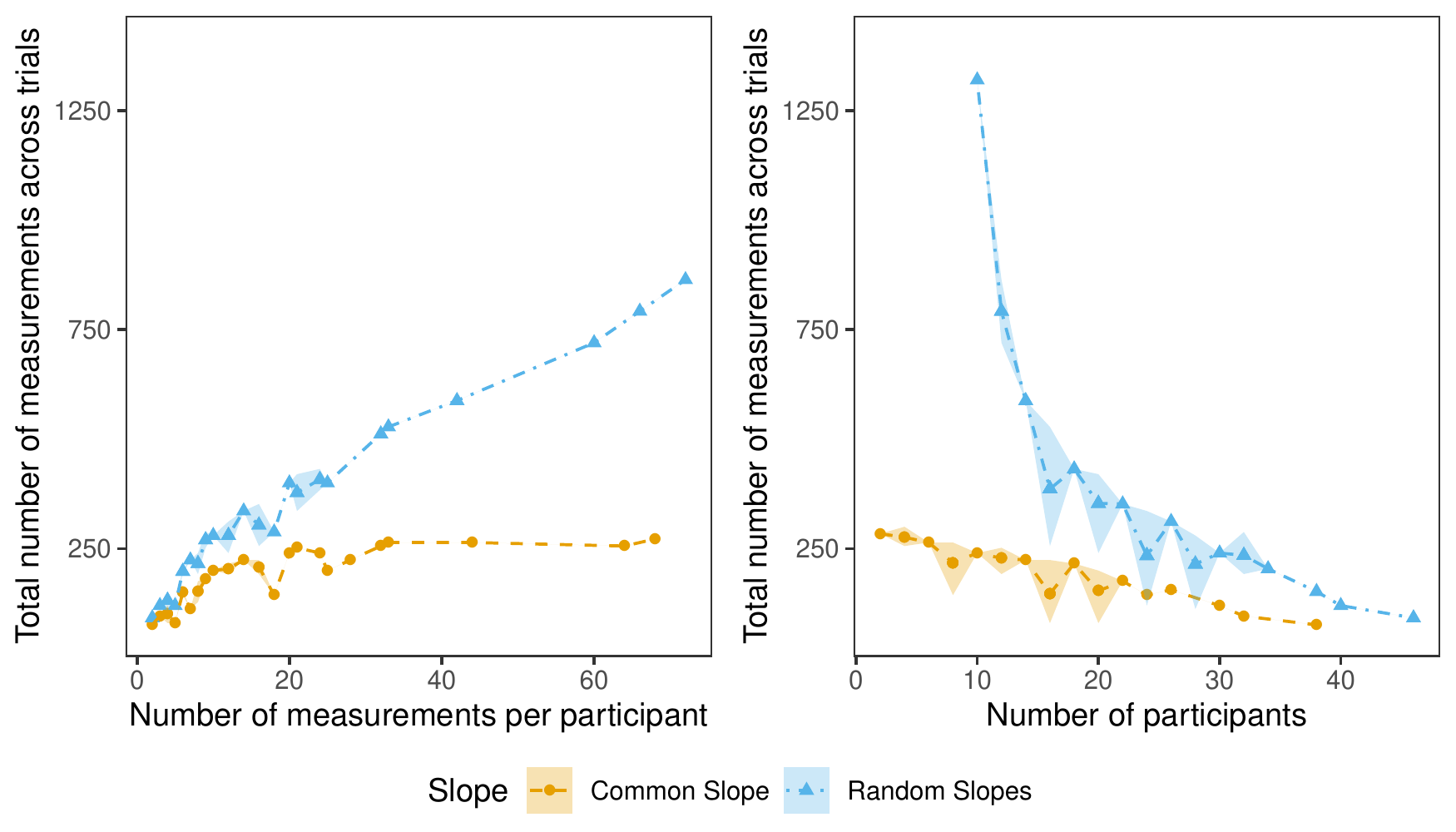}
    \caption{Average required number of measurements across a series of n-of-1 trials versus number of measurements per participant ($KL$, left) and number of participants ($IJ$, right) for optimized designs when fixed-intercept models are used for estimating the population average treatment effect.
    ``Common Slope'' and ``Random Slopes'' refer to ``Fixed-Common'' and ``Fixed-Random'' models in Table \ref{tab: mod_summ_popavg}, respectively. 
    }
    \label{fig: pair_popavg_ar1}
\end{figure}

Figure \ref{fig: pair_popavg_ar1} shows the change in the average required number of measurements across a series of n-of-1 trials ($IJKL$) as a function of the number of measurements per participant ($KL$, left) and the number of participants ($IJ$, right) for optimized designs when fixed-intercept models are used for estimating the population average treatment effect.
Optimized designs refer to designs that achieve criteria with the smallest possible value of the last element with all the other elements in $I$, $J$, $K$ and $L$ fixed. For example, if given the number of sequences $I$, the number of participants assigned to each sequence $J$, and the number of treatment periods in each sequence $K$, a design with $L\geq 5$ measurements in each treatment period will achieve the pre-specified power, the design with $L = 5$ measurements in each treatment period will be the optimized design and will be presented. Analogous optimization applies if $I$, $J$ or $K$ is the last component. The shaded area around the lines represent the range of the required number of measurements across trials from different combinations of $K$ and $L$ (left) and of $I$ and $J$ (right) that lead to the same product and the dots on the lines represent the average if there are multiple combinations. 
As Appendix \ref{app: sim_intcpt_slp} shows that the form of intercepts has almost no effect on the optimized designs for estimating the population average treatment effect, 
we show results from models with fixed intercepts and by the form of slopes.
\cite{legha2018individual} also reported that using restricted maximum likelihood to estimate the population average treatment effect, which gives unbiased estimates of the variances compared with our assumed known variances, the coverage for the effect estimate with fixed or random intercepts are very similar for continuous outcomes. 

In general, the required number of measurements across a series of n-of-1 trials is larger when we use the average of random slopes to estimate the population average treatment effect compared to when using a common slope for estimation because the random slope model contains an extra source of variability from the between-individual variance. 

Additionally, the required number of measurements across trials stays stable with increasing number of measurements per participant when a common slope is used for estimation, while that increases when we allow random effects around the slope; the required number of measurements across trials decreases with increasing number of participants for models with either form of slopes and the decreasing rate is faster when the random-slope model is used for estimation, which leads to the decreasing discrepancy between the required number of measurements across trials for the common- and random-slope model with more participants in trials. These results are consistent with those in \cite{senn2019sample}. There, in a special case of our model described earlier (trials with independent repeated measurements on each participant, one measurement per treatment period and the same number of periods for the two treatments in the sequence), the required number of measurements across trials varied little with the number of measurements per participant for the common effect approach but increased for the random-effect approach. 
The discrepancy between the required number of measurements per participant for the two approaches also decreased with more participants in trials. 

\begin{figure}[htbp]
    \centering
    \includegraphics[width=\textwidth]{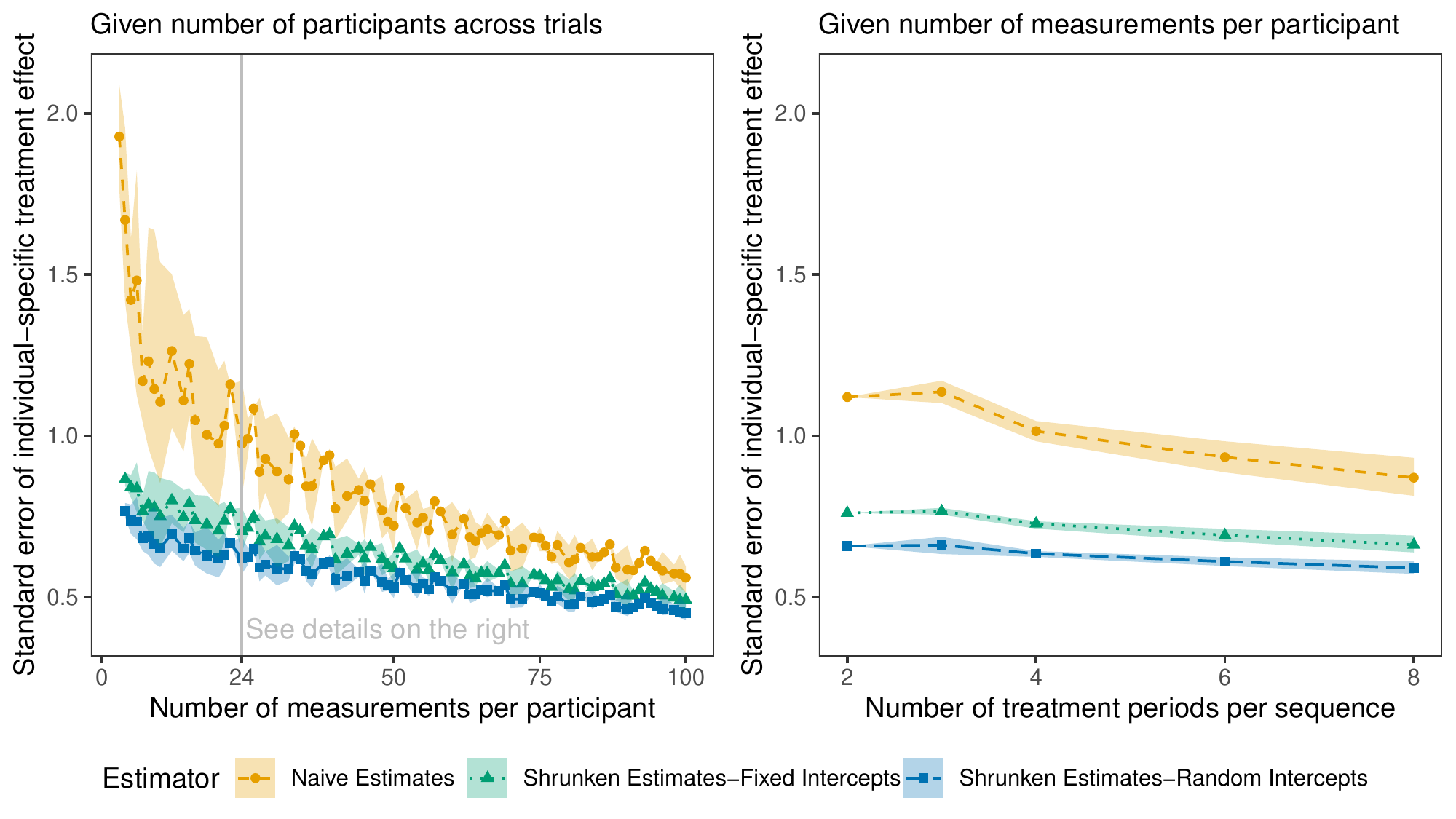}
    \caption{Standard error of naive and shrunken estimates for individual-specific treatment effect versus number of measurements per participant given total number of participants across trials (fixed at 32, left) and versus number of treatment periods per sequence further given number of measurements per participant (fixed at 24, right) for all possible designs that satisfy the power requirement for estimating population average treatment effect.
    ``Naive Estimates'', ``Shrunken Estimates-Fixed Intercepts'' and ``Shrunken Estimates-Random Intercepts'' refer to the standard error of naive estimates, shrunken estimates in the fixed- and random-intercept model, respectively.}
    \label{fig: pair_ind_ar1}
\end{figure}

Figure \ref{fig: pair_ind_ar1} shows the standard errors of the naive and shrunken estimates for individual-specific treatment effects versus the number of measurements per participant given the number of participants (fixed at 32, left)
and versus the number of treatment periods in the sequence further given the number of measurements per participant (fixed at 24, right) for all possible designs that satisfy the power requirement for estimating population average treatment effect.
The shaded areas around the lines represent the range of standard errors from different combinations of $K$ and $L$ that lead to the same product 
(left) and from trials with the same number of treatment periods but with different orders of treatments assigned to the periods (right) and the dots on the lines represent the average if there are multiple combinations.
Note that all possible designs for the series of n-of-1 trials are plotted in the figure because as described in Section \ref{sec: find_design_components} in practice, after using Figure \ref{fig: pair_popavg_ar1} to find designs that satisfy the power requirement for estimating the population average treatment effect, we want to further use Figure \ref{fig: pair_ind_ar1} to finalize the design by picking from all the possible designs that satisfy both the power requirement for estimating population average treatment effect and the standard error requirement for estimating the individual-specific treatment effect. For example, if the required total number of measurements is reasonable when we recruit 32 participants in a series of n-of-1 trials in Figure \ref{fig: pair_popavg_ar1} and we require the standard error of the shrunken estimates with fixed intercepts for individual-specific treatment effects to be lower than 1, all the designs on the ``Shrunken Estimates-Fixed Intercepts'' curve in Figure \ref{fig: pair_ind_ar1} will satisfy the requirement. Additionally, if we are able to measure the outcome 24 times on each participant, Figure \ref{fig: pair_ind_ar1} (right) gives all the possible designs. A specific design can be a series of trials with $I = 4$ possible sequences, $J = 8$ participants assigned to each sequence, $K = 4$ treatment periods per sequence, and $L = 6$ measurements per treatment period. Detailed information for all the possible designs is given in the Shiny app (part (e) in Figure \ref{fig: shiny}).

As expected, the results in Figure \ref{fig: pair_ind_ar1} show the advantage of shrunken estimates over naive estimates for estimating individual-specific treatment effect. Given fixed number of participants, the standard error of naive estimates is larger than that of shrunken estimates. The difference decreases with increasing number of measurements per participant, but even then the standard errors for naive estimates are larger. Naive estimates benefit more from larger number of treatment periods in the sequence when we further fix the number of measurements on each participant, with a larger drop in standard error when we increase the number of treatment periods in the sequence compared to shrunken estimates. 
We also found in Figure \ref{fig: pair_ind_ar1} that the standard error of the shrunken estimates in the random-intercept model is slightly smaller than that in the fixed-intercept model. 

To evaluate the sensitivity of the results to the varying parameters, Appendix \ref{app: add_simulations} includes the following additional illustrations:

\begin{itemize}

\item Appendix \ref{app: sim_alpha_beta} presents results where we use alternative values for the type I and II errors. The average required number of measurements across trials increases with smaller type I and II errors; the rate of increase is higher if 1) we want to reduce smaller type I and II errors, 2) the number of measurements per participant is larger, and 3) the number of participants in trials is smaller. 

\item Appendix \ref{app: sim_Sigma} presents results with alternative parameterizations for the residual error variance matrix. We show results assuming 1) independent and exchangeable correlation structure, and different values for 2) the homogeneous residual error variance and 3) residual correlation coefficient under first order autoregressive correlation structure. 
Trials with exchangeable correlation structure require the fewest measurements across trials, followed by independent correlation structure. Under first order autoregressive correlation structure, the required number of measurements across trials 1) is similar to that under exchangeable correlation structure when the number of measurements per participant is small and the correlation coefficient is large, 2) becomes larger than that under independent correlation structure when the number of measurements per participant is large and the correlation coefficient is small, and 3) increases linearly with homogeneous variance and the rate of increase is higher when the number of measurements per participant is larger and when the number of participants in trials is smaller.

\item Appendix \ref{app: sim_D} presents results with alternative parameterizations for the random-effect variance matrix. Variance of random intercepts and correlation between random intercepts and slopes do not affect the optimized designs that satisfy the power requirement. Holding the number of measurements per participant the same, the average required number of measurements across trials increases linearly with the variance of random slopes; the rate of increase is larger when the number of measurements per participant is larger. Holding the number of participants the same, the average required number of measurements across trials increases more at larger values for the variance of random slopes.

\item Appendix \ref{app: sim_Delta} presents results with alternative minimal clinically important treatment effects. The required number of measurements across trials increases with smaller minimal clinically important treatment effect; the rate of increase is higher if 1) we want to reduce a smaller minimal clinically important treatment effect, 2) the number of measurements per participant is larger, and 3) the number of participants in trials is smaller.

\end{itemize}

\section{Shiny app} \label{sec: shiny}

We implemented our methods in a Shiny app to allow investigators to design n-of-1 trials interactively without requiring programming knowledge or familiarity with a specific software. 
The Shiny app is available at \url{http://jiabeiyang.shinyapps.io/SampleSizeNof1/}. 

\begin{figure}[htbp]
    \centering
    \includegraphics[width=\textwidth]{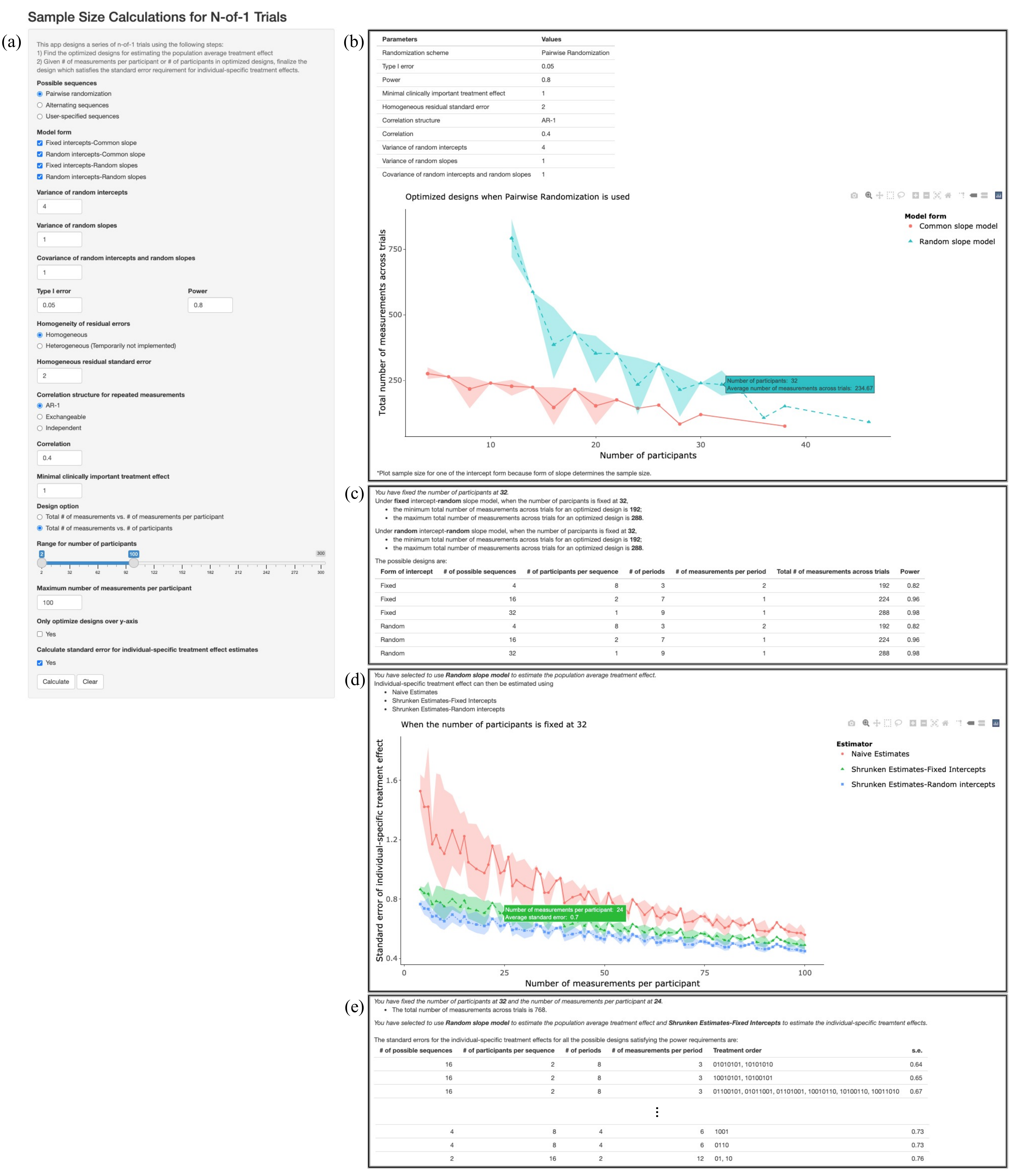}
    \caption{Screenshot of Shiny app. Part (a) allows investigators to specify parameters for designing the series of n-of-1 trials; part (b)-(e) present the possible designs that satisfy the power and standard error requirements specified by the investigators. }
    \label{fig: shiny}
\end{figure}

Figure \ref{fig: shiny} presents the layout of the Shiny app, where we replicated Figure \ref{fig: pair_popavg_ar1} (right) and \ref{fig: pair_ind_ar1}  in Section \ref{sec: simulation}. In Figure \ref{fig: shiny}, part (a) shows the input panel of the app, where investigators are allowed to specify the parameters for designing the series of n-of-1 trials. Part (b)-(e) present information for the possible designs that will satisfy the power and standard error requirements specified by the investigators. Part (b) presents the specified parameters in the input panel and how the average required number of measurements across trials changes as a function of the number of participants for optimized designs (Figure \ref{fig: pair_popavg_ar1}, right). Part (c) and (d) show the detailed design information for optimized designs and the standard errors of the individual-specific treatment effect estimates for all possible designs that satisfy the power requirement (Figure \ref{fig: pair_ind_ar1}), respectively, when one fixes the number of participants across trials by clicking on a specific point in the figure in part (b). Part (e) shows the detailed information for all possible designs that satisfy the power and standard error requirements when one further fixes the number of measurements per participant by clicking on a point in the figure in part (d). If investigators want to know how the average required number of measurements across trials changes as a function of the number of measurements per participant (Figure \ref{fig: pair_popavg_ar1}, left), they can choose ``Total \# of measurements vs. \# of measurements per participant" under "Design option'' in the input panel. The output in part (d) of the Shiny app will be replaced accordingly by presenting the standard errors versus the number of participants in the trials given the number of measurements per participant.

The Shiny app implements both alternating sequences and pairwise randomization and also allows the user to upload manually specified sequences through ``Possible sequences''-``User-specified sequences'' in the input panel.
Because the number of possible sequences under restricted and unrestricted randomization becomes large as the number of treatment periods in the sequences increases (Table \ref{tab: numseq_k}) and many sequences may be impractical if the same treatment is given in too many consecutive treatment periods,
users can complete calculations for these two randomization schemes by picking specific sequences of interest and uploading them through ``User-specified sequences''. 

Additionally, we provide an option to only optimize designs over the scale of the y-axis in part (b) of the Shiny app. Because of the current optimization, even if we allow large number of participants in the trials, the maximum number of participants presented in part (b) of the Shiny app will be small because designs with more participants are not optimized using our definition of optimized designs. Therefore, this option will present all the possible designs within the specified range for the x-axis, only optimized on the scale of the y-axis. Finally, if investigators are not interested in estimating the individual-specific treatment effects, they can clear ``Calculate standard error for individual-specific treatment effect estimates'' and only part (b) and (c) will be displayed in the output.

\section{Discussion} \label{sec: discussion}

We present a procedure for calculating the sample size for n-of-1 trials. We formally define the design components for determining the sample size of a series of n-of-1 trials which include the sequences with different orders of treatments assigned to periods in the sequences, the number of participants assigned to each sequence, 
and the number of measurements in each treatment period. We present models for analyzing n-of-1 trials and use them to derive the required sample size for estimating population average treatment effect and the standard error of individual-specific treatment effect estimates. 
We recommend that investigators first use the sample size formula to find the possible combinations of the design components that will satisfy the power requirement for estimating the population average treatment effect, and, if of interest, use the standard error formulae to pick the combinations of design components that will also satisfy the standard error requirements for the individual-specific treatment effect estimates.
We implement and illustrate the procedure in the paper and through a Shiny app.

Several directions of future research are practically useful. First, the current derivations assume that the variance components, either for residual errors or for random effects, are known or estimated with adequate precision. Taking into account the fact that the variance components are estimated will involve using a distribution instead of the standard normal distribution when deriving power in Section \ref{sec: power_popavg} and will facilitate more accurate planning of the n-of-1 trials. Second, it will be helpful to extend the results to discrete outcomes, although correlation among the repeated measurements would then need to be handled differently \citep{zeger1988regression}. Additionally, it will be valuable to extend the results to trials with more than two treatments \citep{kravitz2020feasibility} and with multiple outcomes of interest \citep{barr2015preempt}. Lastly, 
to avoid carryover effects, we can introduce washout periods into the derivations when switching from one treatment to another. These periods will limit the number of switches between different treatments in the sequences.

\section*{Code}
\label{sec: supp}

The R code for illustrations and the Shiny app and an example input file for the Shiny app are available on \url{http://github.com/jiabei-yang/SampleSizeNof1}. The Shiny app is hosted at \url{http://jiabeiyang.shinyapps.io/SampleSizeNof1/}.


\newpage 

\FloatBarrier

\section*{{\bf  Supplementary Web Appendix}}

\setcounter{section}{0}
\renewcommand{\thesection}{S.\arabic{section}}
\renewcommand{\theHsection}{Supplement.\thesection}

\setcounter{equation}{0}
\renewcommand{\theequation}{S-\arabic{equation}}
\renewcommand{\theHequation}{Supplement.\theequation}

\setcounter{table}{0}
\renewcommand{\thetable}{S-\arabic{table}}
\renewcommand{\theHtable}{Supplement.\thetable}

\setcounter{figure}{0}
\renewcommand{\thefigure}{S-\arabic{figure}}
\renewcommand{\theHfigure}{Supplement.\thefigure}

\setcounter{page}{1}
\renewcommand{\thepage}{\arabic{page}}

References to figures, tables, theorems and equations preceded by ``S-'' are internal to this supplement; all other references refer to the main paper.

\section{Derivations for the variance of the shrunken estimates} \label{app: prec_shrunk}

Write the variance as
\begin{align} \label{eq: var_single_shrunk}
    \text{Var}(\hat{\delta}_{i^*j^*} - \delta_{i^*j^*}) =& \text{Var}\left[(\hat{\delta} + \hat{\gamma}_{\delta i^*j^*}) - (\delta + \gamma_{\delta i^*j^*}) \right]\nonumber\\
    =& \text{Var}(\hat{\delta}) + \text{Var}(\hat{\gamma}_{\delta i^*j^*} - \gamma_{\delta i^*j^*}) + 2 \text{cov}(\hat{\delta}, \hat{\gamma}_{\delta i^*j^*} - \gamma_{\delta i^*j^*})\nonumber\\
    =& \bm{C}_{\bm{\theta}}\text{Var}(\hat{\bm{\theta}})\bm{C}_{\bm{\theta}}^T + \bm{C}_{\bm{b}}\text{Var}(\hat{\bm{b}}_{i^*j^*} - \bm{b}_{i^*j^*})\bm{C}_{\bm{b}}^T + 2\bm{C}_{\bm{\theta}} \text{cov}(\hat{\bm{\theta}}, \hat{\bm{b}}_{i^*j^*}) \bm{C}_{\bm{b}}^T - 2\bm{C}_{\bm{\theta}} \text{cov}(\hat{\bm{\theta}}, \bm{b}_{i^*j^*}) \bm{C}_{\bm{b}}^T
\end{align}
where $\bm{C}_{\bm{\theta}}$ are summarized in Table \ref{tab: mod_summ_popavg}; $\bm{C}_{\bm{b}}$ is the contrast matrix that pulls off the appropriate element from the random effects vector or the variance matrix for the random effects and equals $1$ for the model with fixed intercepts and random slopes and equals $(0, 1)$ for the model with random intercepts and random slopes.

Using $\hat{\bm{\theta}} = \left[ \sum_{i = 1}^I \sum_{j = 1}^{J_i} \mathbf{X}_{ij}^T \mathbf{\Sigma}_{ij}^{-1} \mathbf{X}_{ij} \right]^{-1} \left[ \sum_{i = 1}^I \sum_{j = 1}^{J_i} \mathbf{X}_{ij}^T \mathbf{\Sigma}_{ij}^{-1} \bm{Y}_{ij} \right]$ and $\hat{\bm{b}}_{i^*j^*} = \mathbf{D}\mathbf{Z}_{i^*j^*}^T \mathbf{\Sigma}_{i^*j^*}^{-1}(\bm{Y}_{i^*j^*} - \mathbf{X}_{i^*j^*} \hat{\bm{\theta}})$ \citep{laird1982random},
\begin{align} \label{eq: cov_popavg_estfixestrand}
    \text{cov}(\hat{\bm{\theta}}, \hat{\bm{b}}_{i^*j^*}) 
    =& \text{cov}\left\{\hat{\bm{\theta}},  \mathbf{D}\mathbf{Z}_{i^*j^*}^T \mathbf{\Sigma}_{i^*j^*}^{-1} \bm{Y}_{i^*j^*} \right\} - \text{cov}\left\{ \hat{\bm{\theta}}, \mathbf{D}\mathbf{Z}_{i^*j^*}^T \mathbf{\Sigma}_{i^*j^*}^{-1} \mathbf{X}_{i^*j^*} \hat{\bm{\theta}} \right\} \nonumber\\ 
    =& \text{cov}\left\{ \left[ \sum_{i = 1}^I \sum_{j = 1}^{J_i} \mathbf{X}_{ij}^T \mathbf{\Sigma}_{ij}^{-1} \mathbf{X}_{ij} \right]^{-1} \mathbf{X}_{i^*j^*}^T \mathbf{\Sigma}_{i^*j^*}^{-1} \bm{Y}_{i^*j^*},  \mathbf{D}\mathbf{Z}_{i^*j^*}^T \mathbf{\Sigma}_{i^*j^*}^{-1} \bm{Y}_{i^*j^*} \right\} -\text{Var}( \hat{\bm{\theta}}) \mathbf{X}_{i^*j^*}^T \mathbf{\Sigma}_{i^*j^*}^{-1} \mathbf{Z}_{i^*j^*} \mathbf{D} \nonumber\\
    & \qquad\qquad \text{(participants are independent of each other)} \nonumber\\
    =& 0
\end{align}
\begin{align} \label{eq: cov_popavg_estfixrand}
    \text{cov}(\hat{\bm{\theta}}, \bm{b}_{i^*j^*}) 
    =& \text{cov}\left\{ \left[ \sum_{i = 1}^I \sum_{j = 1}^{J_i} \mathbf{X}_{ij}^T \mathbf{\Sigma}_{ij}^{-1} \mathbf{X}_{ij} \right]^{-1} \mathbf{X}_{i^*j^*}^T \mathbf{\Sigma}_{i^*j^*}^{-1} \bm{Y}_{i^*j^*}, \bm{b}_{i^*j^*} \right\} \nonumber\\ 
    =& \text{cov}\left\{ \left[ \sum_{i = 1}^I \sum_{j = 1}^{J_i} \mathbf{X}_{ij}^T \mathbf{\Sigma}_{ij}^{-1} \mathbf{X}_{ij} \right]^{-1} \mathbf{X}_{i^*j^*}^T \mathbf{\Sigma}_{i^*j^*}^{-1} \mathbf{Z}_{i^*j^*}\bm{b}_{i^*j^*},  \bm{b}_{i^*j^*} \right\} \nonumber\\
    & \qquad\qquad (\bm{b}_{i^*j^*} \independent \bm{\epsilon}_{i^*j^*}) \nonumber\\
    =& \left[ \sum_{i = 1}^I \sum_{j = 1}^{J_i} \mathbf{X}_{ij}^T \mathbf{\Sigma}_{ij}^{-1} \mathbf{X}_{ij} \right]^{-1} \mathbf{X}_{i^*j^*}^T \mathbf{\Sigma}_{i^*j^*}^{-1} \mathbf{Z}_{i^*j^*} \mathbf{D}. 
\end{align}
Plug $\text{Var}(\hat{\bm{\theta}}) = \left( \sum_{i=1}^I \sum_{j = 1}^{J_i} \mathbf{X}_{ij}^T \mathbf{\Sigma}_{ij}^{-1} \mathbf{X}_{ij} \right)^{-1}$,\\
$\text{Var}(\hat{\bm{b}}_{i^*j^*} - \bm{b}_{i^*j^*}) = \mathbf{D} - \mathbf{D} \mathbf{Z}_{i^*j^*}^T \mathbf{\Sigma}_{i^*j^*}^{-1} \mathbf{Z}_{i^*j^*} \mathbf{D} + \mathbf{D} \mathbf{Z}_{i^*j^*}^T \mathbf{\Sigma}_{i^*j^*}^{-1} \mathbf{X}_{i^*j^*}\left( \sum_{i=1}^I \sum_{j = 1}^{J_i} \mathbf{X}_{ij}^T \mathbf{\Sigma}_{ij}^{-1} \mathbf{X}_{ij} \right)^{-1} \mathbf{X}_{i^*j^*}^T \mathbf{\Sigma}_{i^*j^*}^{-1} \mathbf{Z}_{i^*j^*} \mathbf{D}$ \citep{laird1982random}, \eqref{eq: cov_popavg_estfixestrand} and \eqref{eq: cov_popavg_estfixrand} into \eqref{eq: var_single_shrunk} gives $\text{Var}(\hat{\delta}_{i^*j^*} - \delta_{i^*j^*})$.

\section{Additional comparisons of designs} \label{app: add_simulations}

We include additional illustrations in this section. Unless specified, the parameter choices were the same as those described in Section \ref{sec: simulation}.

\subsection{Results for models with fixed and random intercepts} \label{app: sim_intcpt_slp}

\begin{figure}[htbp]
    \centering
    \includegraphics[width=\textwidth]{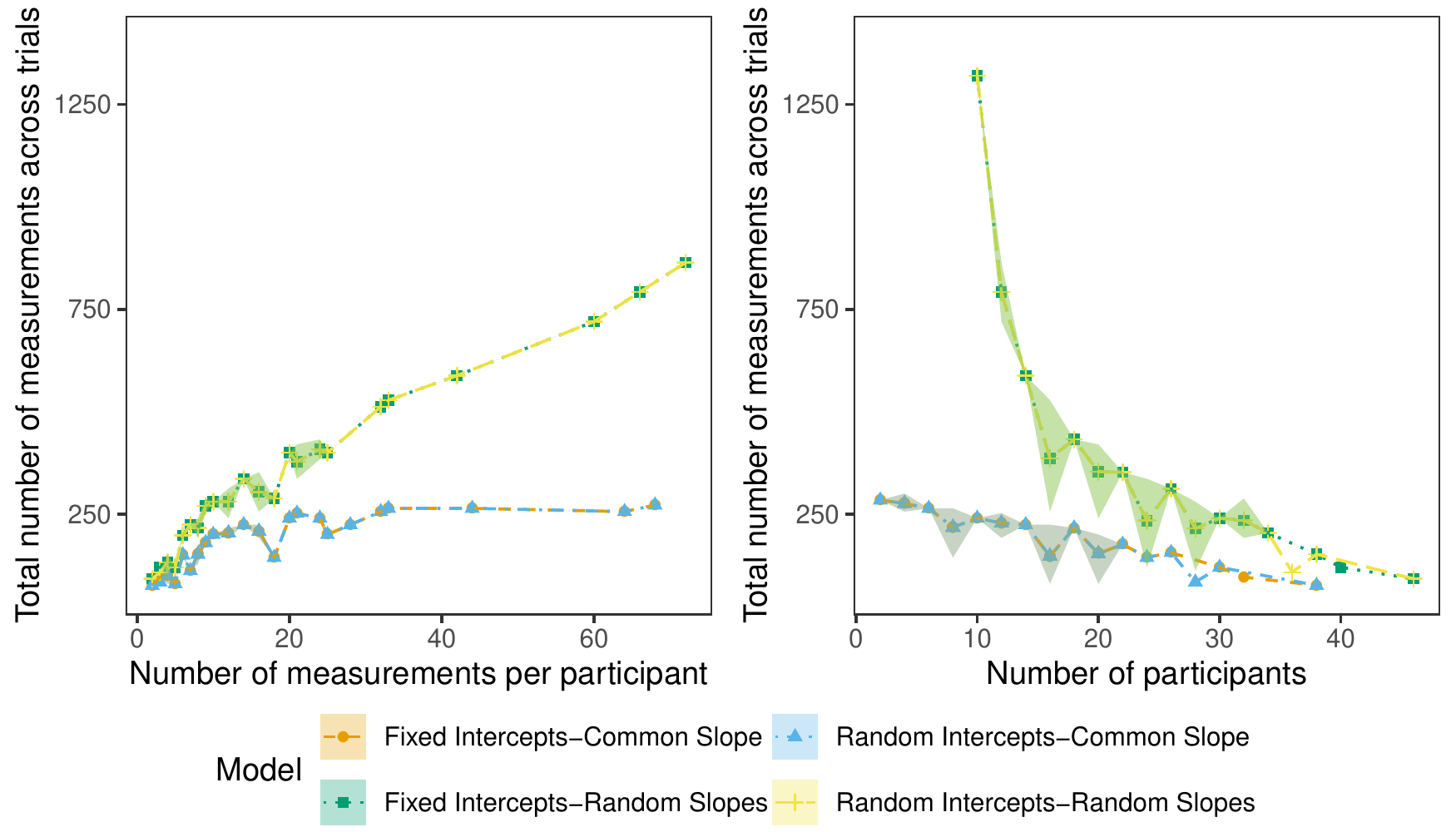}
    \caption{Average required number of measurements across a series of n-of-1 trials versus number of measurements per participant ($KL$, left) and number of participants ($IJ$, right) for optimized designs 
    when the four possible models in Table \ref{tab: mod_summ_popavg} are used for estimating the population average treatment effect.
    }
    \label{fig: pair_popavg_ar1_intcpt_slp}
\end{figure}

Figure \ref{fig: pair_popavg_ar1_intcpt_slp} shows how the average required number of measurements across a series of n-of-1 trials ($IJKL$) changes as a function of the number of measurements per participant ($KL$, left) or the number of participants ($IJ$, right) for optimized designs when the four possible models in Table \ref{tab: mod_summ_popavg} are used for estimating the population average treatment effect.

We find in Figure \ref{fig: pair_popavg_ar1_intcpt_slp} that the optimized designs for estimating the population average treatment effect are similar if we choose to model the slopes in the same way 
regardless of how we choose to model the intercepts.

Therefore in Section \ref{sec: simulation} and the following sections, we will present results for models with fixed intercepts and by the form of slopes.

\subsection{Results with alternative type I and II errors} \label{app: sim_alpha_beta}

\begin{figure}[htbp]
    \centering
    \includegraphics[width=\textwidth]{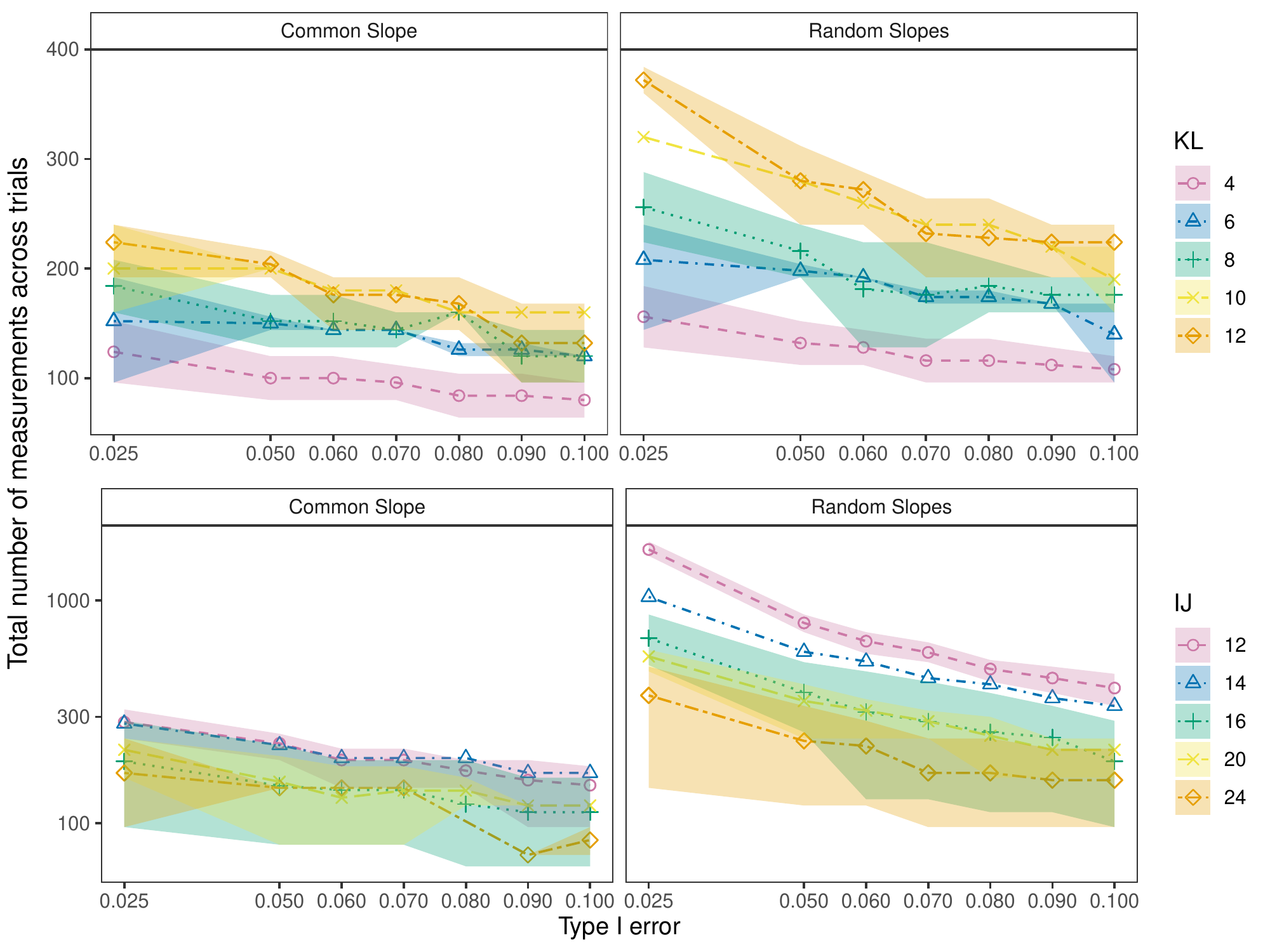}
    \caption{Average required number of measurements across a series of n-of-1 trials versus type I error rates for optimized designs when we use a common slope (left) and the average of random slopes (right) to estimate the population average treatment effect fixing the number of measurements per participant ($KL$, top) and the number of participants ($IJ$, bottom).
    }
    \label{fig: pair_popavg_alpha}
\end{figure}

\begin{figure}[htbp]
    \centering
    \includegraphics[width=\textwidth]{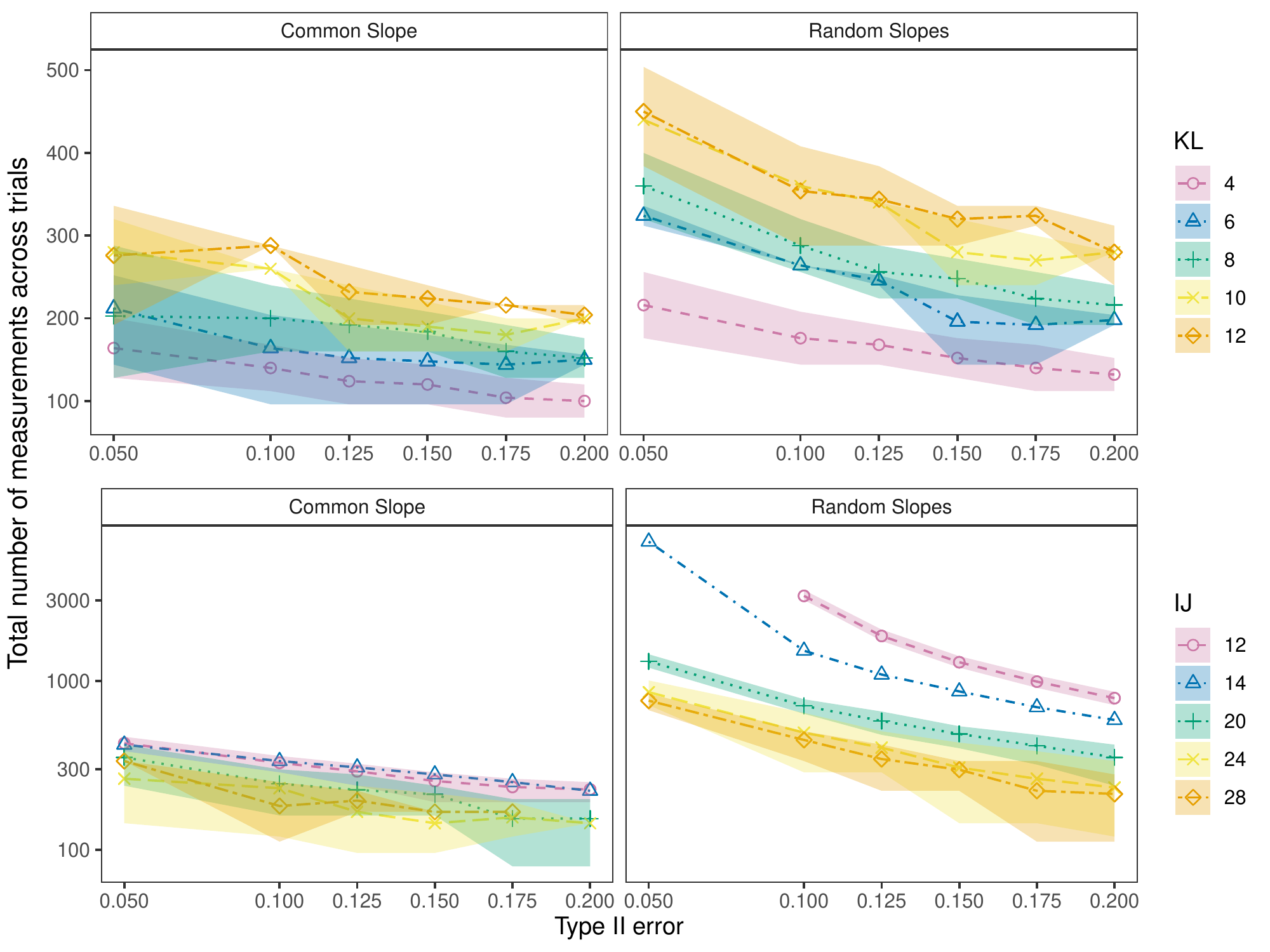}
    \caption{Average required number of measurements across a series of n-of-1 trials versus type II error rates for optimized designs when we use a common slope (left) and the average of random slopes (right) to estimate the population average treatment effect fixing the number of measurements per participant ($KL$, top) and the number of participants ($IJ$, bottom).
    }
    \label{fig: pair_popavg_beta}
\end{figure}

Figure \ref{fig: pair_popavg_alpha} and \ref{fig: pair_popavg_beta} show how the average required number of measurements across a series of n-of-1 trials ($IJKL$) changes as a function of type I and II errors, respectively, for optimized designs when we use a common slope (left) and the average of random slopes (right) to estimate the population average treatment effect fixing the number of measurements per participant ($KL$, top) and the number of participants ($IJ$, bottom). Other parameter choices were the same as those described in Section \ref{sec: simulation}.

As expected, the results show that the average required number of measurements across trials increases with smaller type I and II errors. The increasing rate is higher 1) if we want to reduce smaller type I and II errors, 2) if the number of measurements per participant is larger and 3) if the number of participants in trials is smaller.

\subsection{Results with alternative parameterizations for residual error variance matrix} \label{app: sim_Sigma}

\begin{figure}[htbp]
    \centering
    \includegraphics[width=\textwidth]{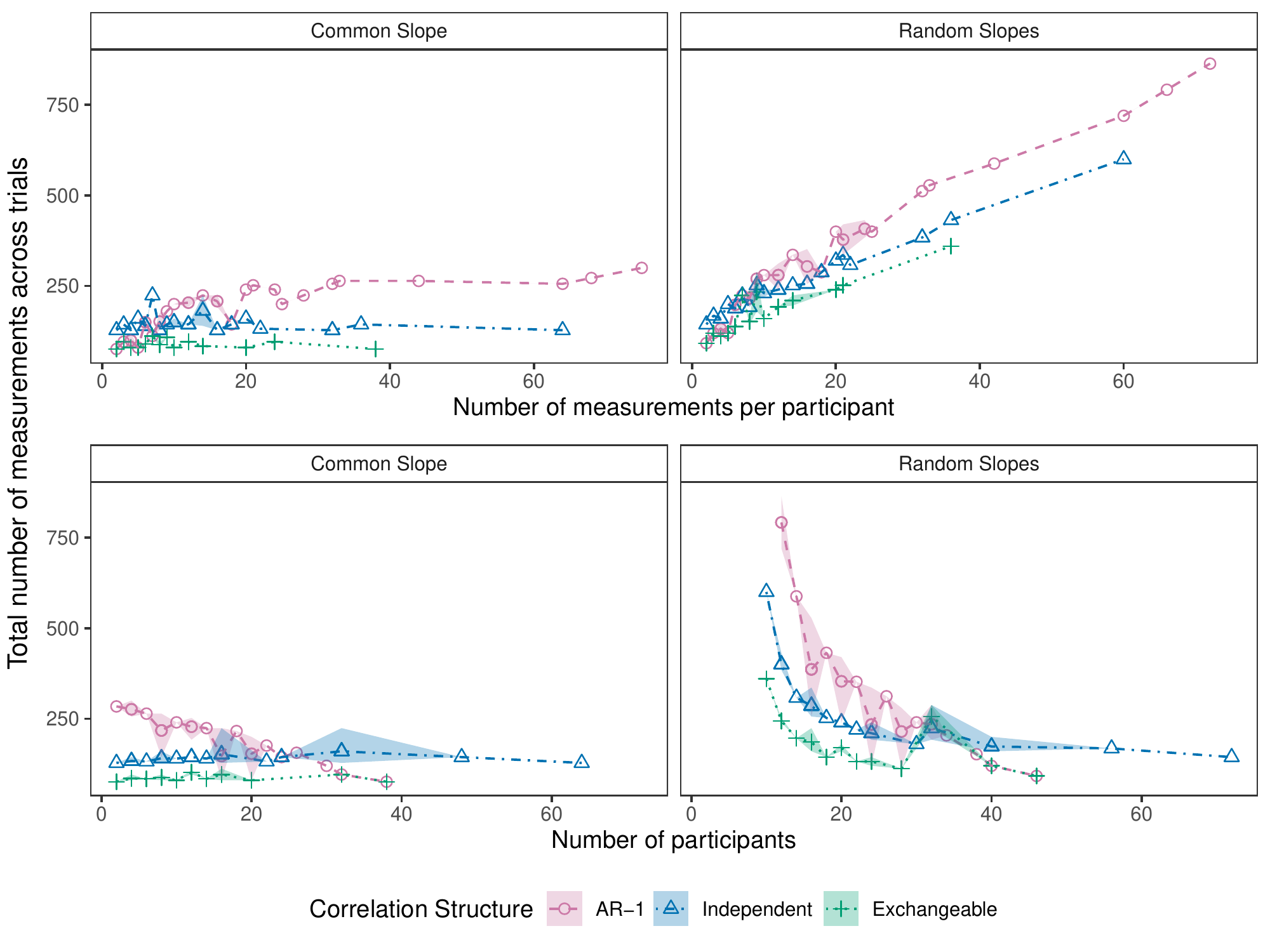}
    \caption{Average required number of measurements across a series of n-of-1 trials versus number of measurements per participant ($KL$, top) and the number of participants ($IJ$, bottom) for optimized designs assuming independent, exchangeable and first order autoregressive correlation structures among the repeated measurements on a single participant when we use a common slope (left) and the average of random slopes (right) to estimate the population average treatment effect.
    }
    \label{fig: resid_corrstr}
\end{figure}

\begin{figure}[htbp]
    \centering
    \includegraphics[width=\textwidth]{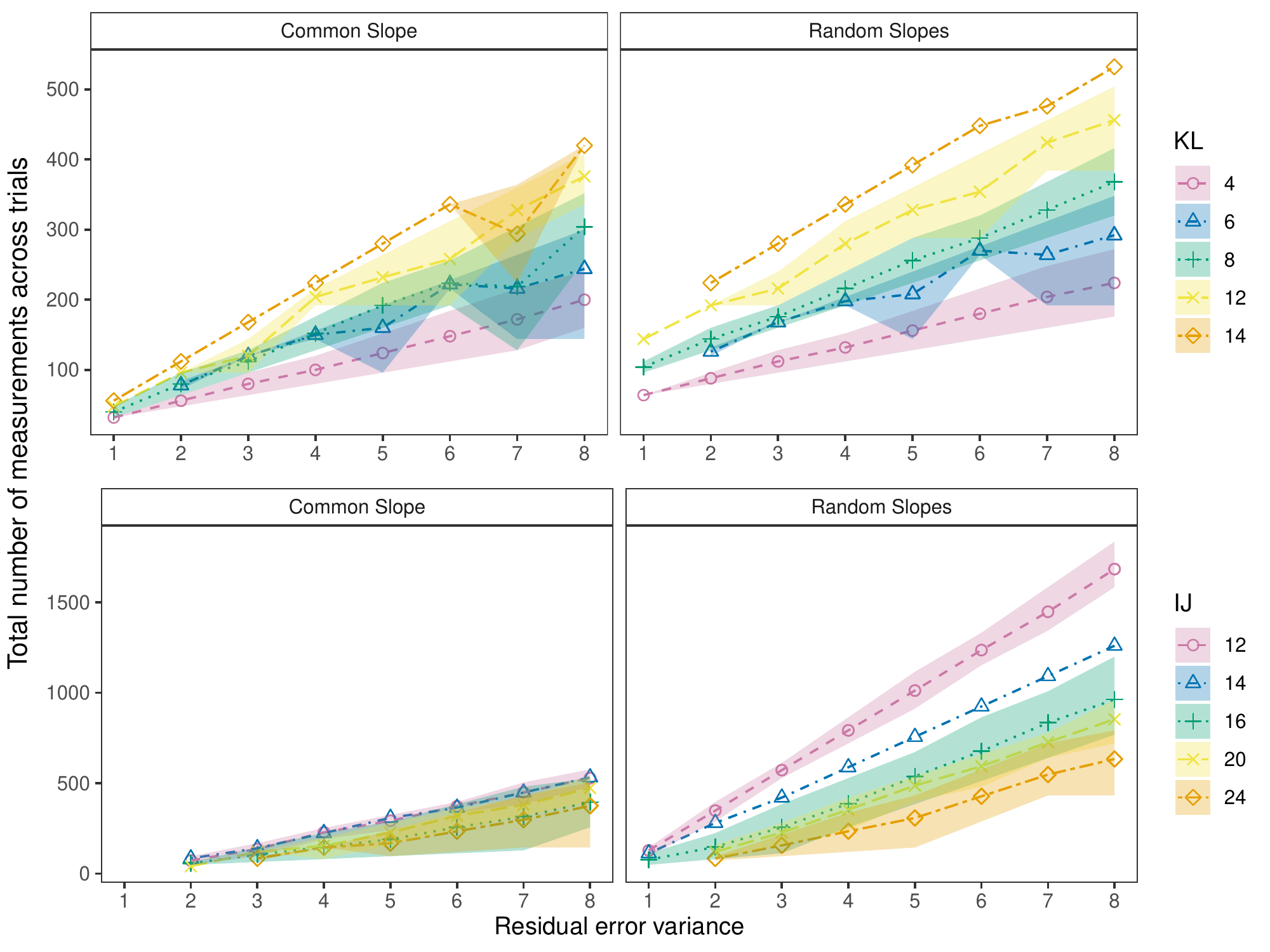}
    \caption{Average required number of measurements across a series of n-of-1 trials versus residual error variance for optimized designs when we use a common slope (left) and the average of random slopes (right) to estimate the population average treatment effect fixing the number of measurements per participant ($KL$, top) and the number of participants ($IJ$, bottom). 
    }
    \label{fig: resid_sgmepsl}
\end{figure}

\begin{figure}[htbp]
    \centering
    \includegraphics[width=\textwidth]{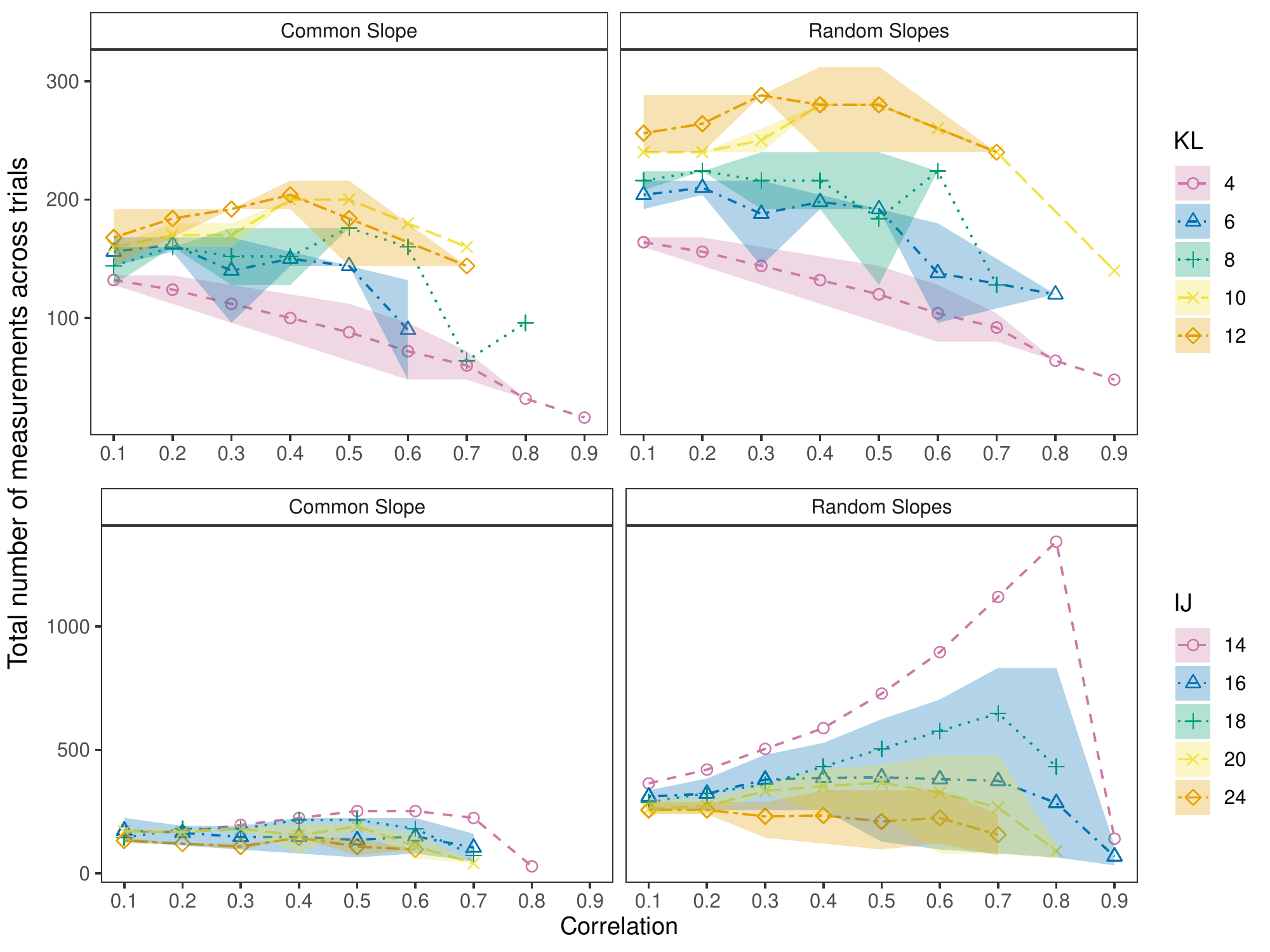}
    \caption{Average required number of measurements across a series of n-of-1 trials versus residual correlation coefficient for optimized designs when we use a common slope (left) and average of random slopes (right) to estimate the population average treatment effect fixing the number of measurements per participant ($KL$, top) and the number of participants ($IJ$, bottom).
    }
    \label{fig: resid_rho}
\end{figure}

Figure \ref{fig: resid_corrstr} shows how the average required number of measurements across a series of n-of-1 trials ($IJKL$) changes as a function of the number of measurements per participant ($KL$, top) and the number of participants ($IJ$, bottom) for optimized designs assuming independent, exchangeable and first order autoregressive correlation structures among the repeated measurements on a single participant when we use a common slope (left) and the average of random slopes (right) to estimate the population average treatment effect. Figure \ref{fig: resid_sgmepsl} and \ref{fig: resid_rho} show how the average required number of measurements across trials ($IJKL$) changes as a function of the residual error variance and the residual correlation coefficient, respectively, fixing the number of measurements per participant ($KL$, top) and the number of participants ($IJ$, bottom) for optimized designs. Other parameter choices were the same as those described in Section \ref{sec: simulation}.

Trials with exchangeable correlation structure require the fewest measurements across trials, followed by independent correlation structure. Under first order autoregressive correlation structure, the required number of measurements across trials 1) is similar to that under exchangeable correlation structure when the number of measurements per participant is small and the correlation coefficient is large and 2) becomes larger than that under independent correlation structure when the number of measurements per participant is large and the correlation coefficient is small, and 3) increases linearly with homogeneous variance and the increasing rate is larger when the number of measurements per participant is larger and when the number of participants in trials is smaller.

\subsection{Results with alternative parameterizations for random-effect variance matrix} \label{app: sim_D}

\begin{figure}[htbp]
    \centering
    \includegraphics[width=\textwidth]{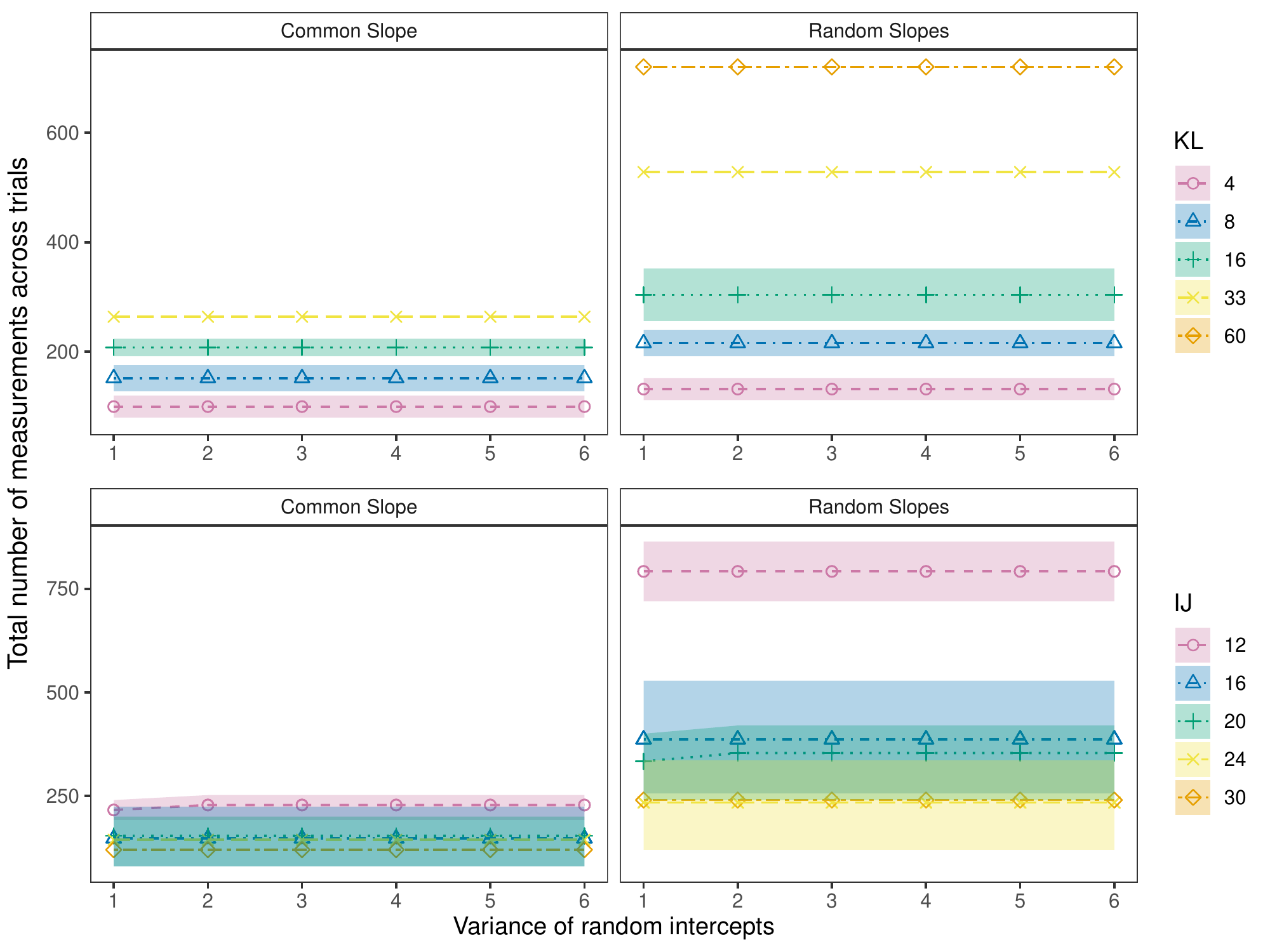}
    \caption{Average required number of measurements across a series of n-of-1 trials versus variance of random intercepts when we use random intercepts-common slope model (left) and random intercepts-random slopes model (right) to estimate the population average treatment effect fixing the number of measurements per participant ($KL$, top) and the number of participants ($IJ$, bottom) for optimized designs. 
    }
    \label{fig: randeff_sgmMu}
\end{figure}

\begin{figure}[htbp]
    \centering
    \includegraphics[width=\textwidth]{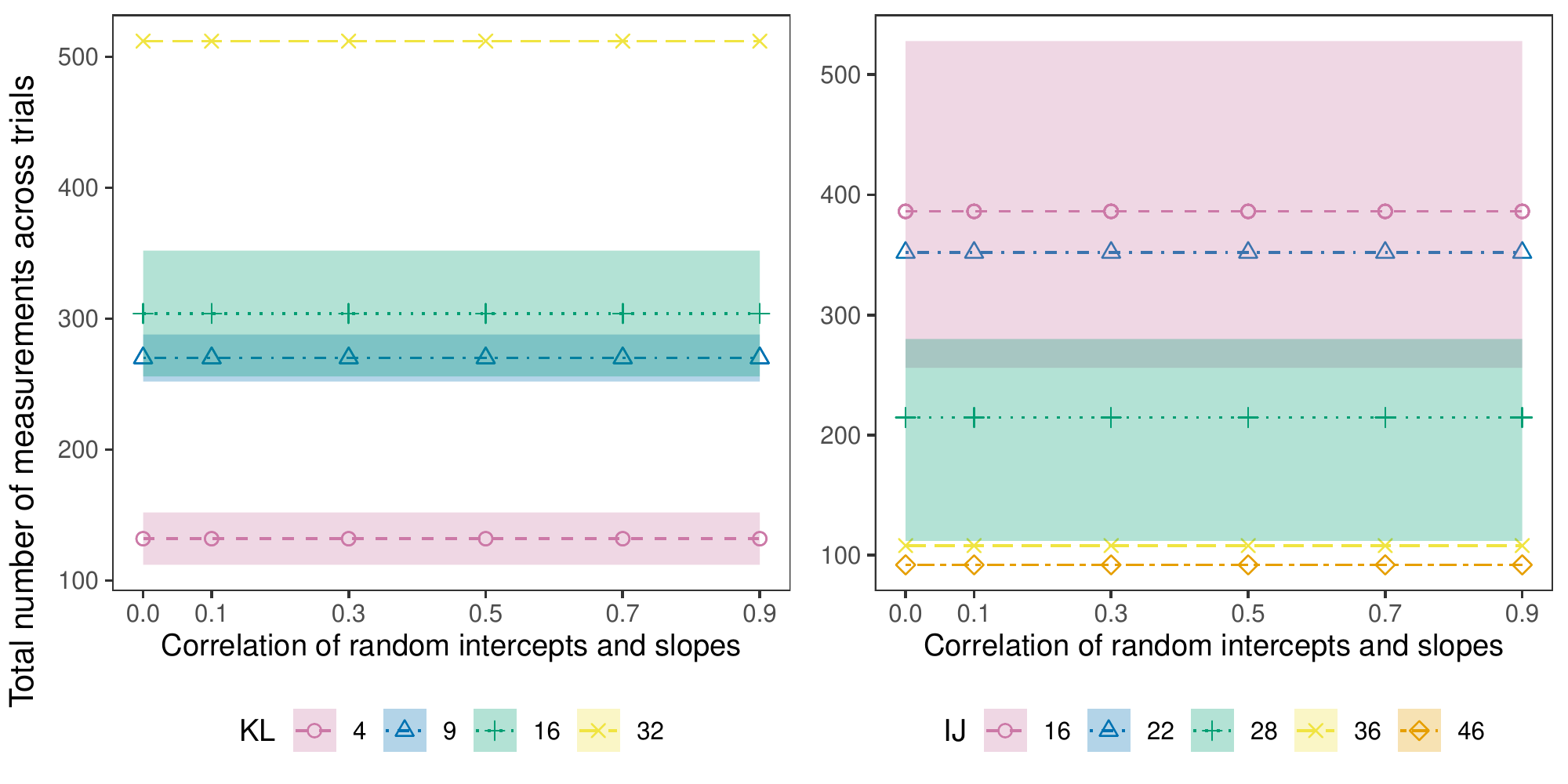}
    \caption{Average required number of measurements across a series of n-of-1 trials versus correlation between random intercepts and random slopes when we use random intercepts-random slopes model to estimate the population average treatment effect fixing the number of measurements per participant ($KL$, left) and the number of participants ($IJ$, right) for optimized designs.
    }
    \label{fig: randeff_sgmMuTau}
\end{figure} 

\begin{figure}[htbp]
    \centering
    \includegraphics[width=\textwidth]{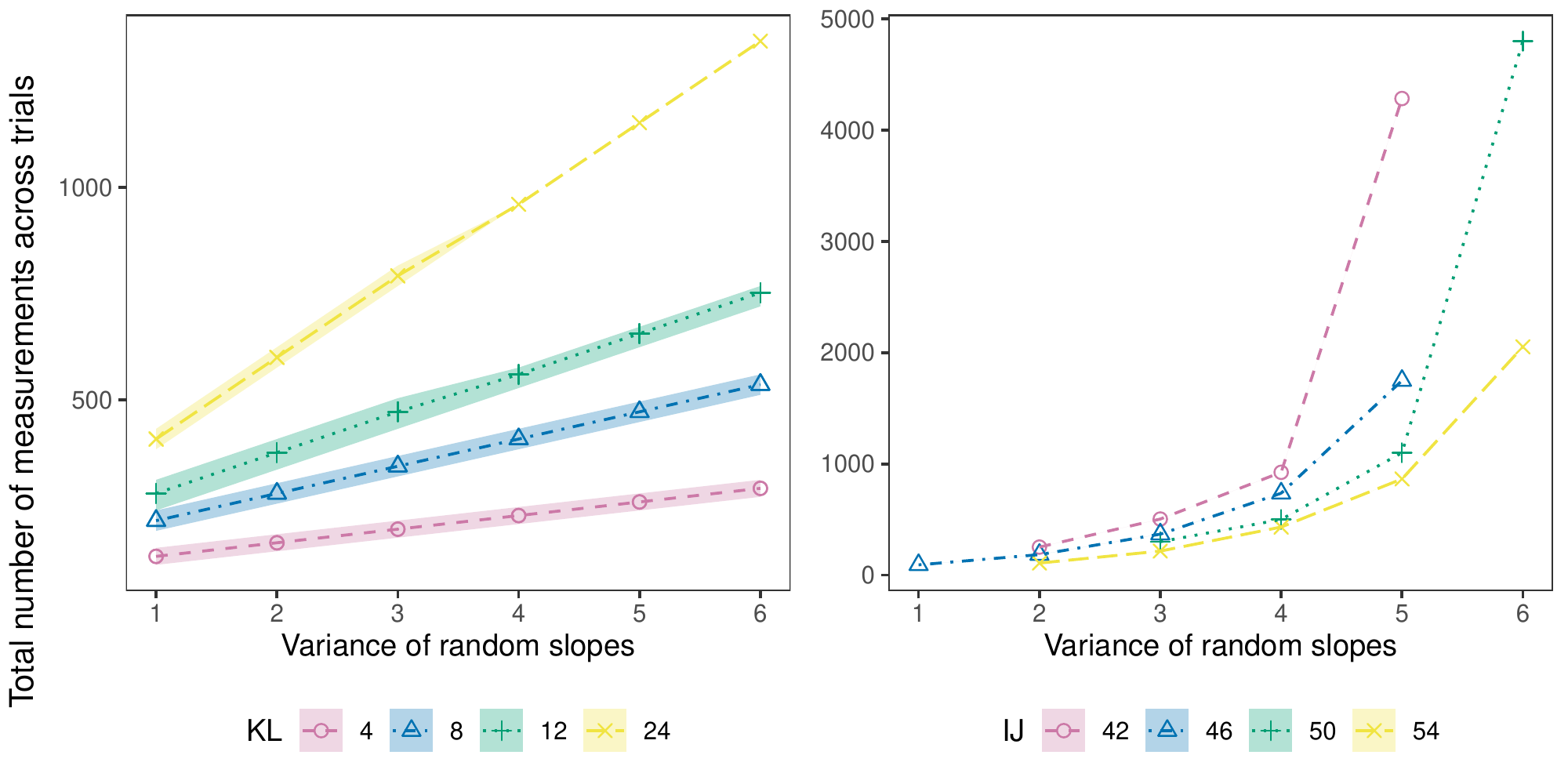}
    \caption{Average required number of measurements across a series of n-of-1 trials versus variance of random slopes when we use the average of random slopes to estimate the population average treatment effect fixing the number of measurements per participant ($KL$, left) and the number of participants ($IJ$, right) for optimized designs. 
    }
    \label{fig: randeff_sgmTau}
\end{figure} 

Figure \ref{fig: randeff_sgmMu}, \ref{fig: randeff_sgmMuTau} and \ref{fig: randeff_sgmTau} show how the average required number of measurements across a series of n-of-1 trials changes as a function of the variance of random intercepts, correlation between random intercepts and slopes and the variance of random slopes, respectively, when applicable in the model for optimized designs.

The results show that variance of random intercepts and correlation between random intercepts and slopes do not affect the optimized designs that satisfy the power requirement. Holding the number of measurements per participant the same, the average required number of measurements across trials increases linearly with the variance of random slopes; the increasing rate is larger when the number of measurements per participant is larger. Holding the number of participants the same, the average required number of measurements across trials increases more at larger values of variance of random slopes.

\subsection{Results with alternative minimal clinically important treatment effects} \label{app: sim_Delta}

\begin{figure}[htbp]
    \centering
    \includegraphics[width=\textwidth]{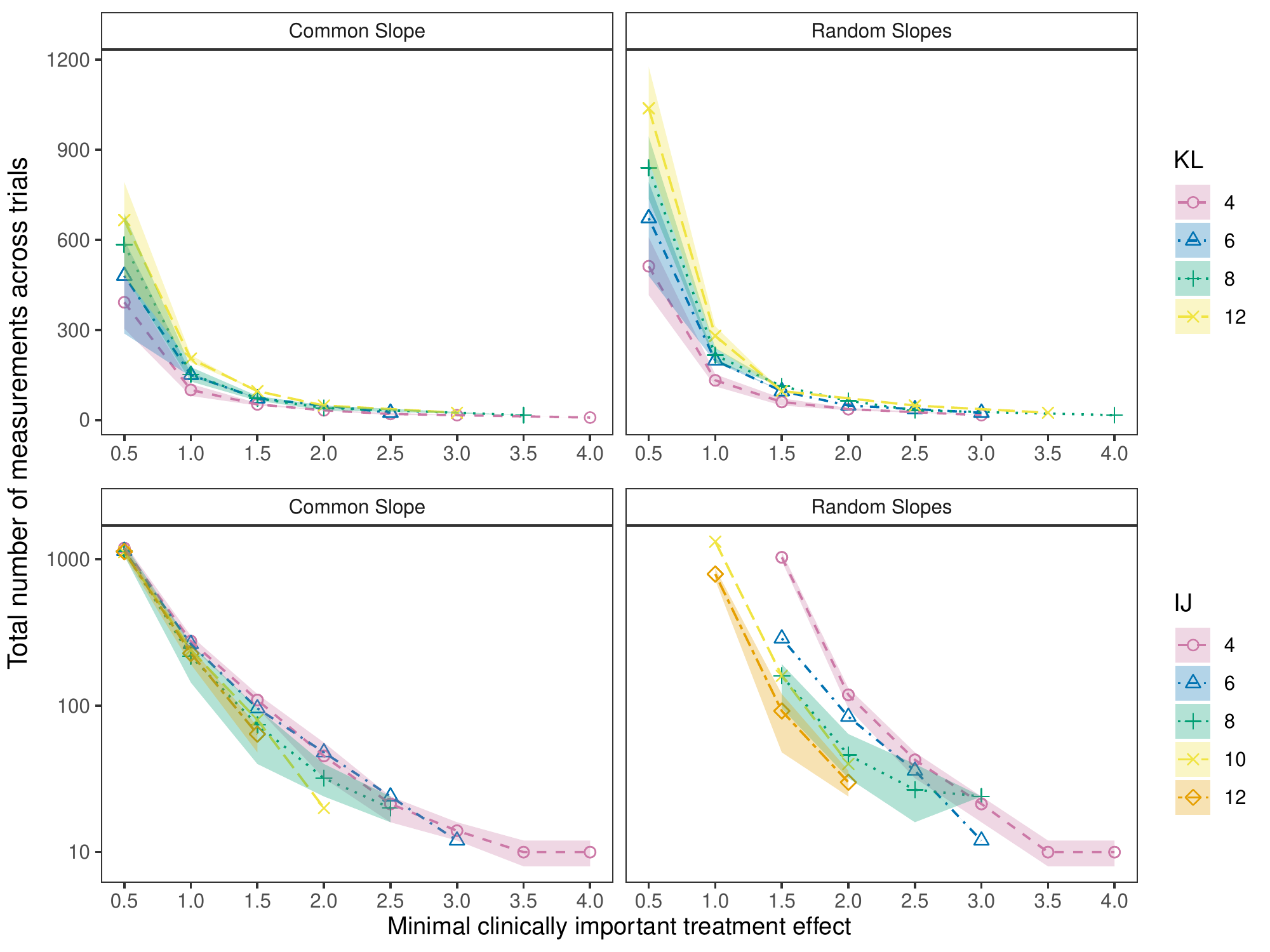} 
    \caption{Average required number of measurements across a series of n-of-1 trials versus minimal clinically important treatment effect when we use a common slope (left) and the average of random slopes (right) to estimate the population average treatment effect fixing the number of measurements per participant ($KL$, top) and the number of participants ($IJ$, bottom) for optimized designs
    }
    \label{fig: delta}
\end{figure}

Figure \ref{fig: delta} shows how the average required number of measurements across a series of n-of-1 trials ($IJKL$) changes as a function of minimal clinically important treatment effect when we use a common slope (left) and the average of random slopes (right) to estimate the population average treatment effect  fixing the number of measurements per participant ($KL$, top) and the number of participants ($IJ$, bottom) for optimized designs.

The results show that the required number of measurements across trials increases with smaller minimal clinically important treatment effect. The increasing rate is higher 1) if we want to reduce a smaller minimal clinically important treatment effect, 2) if the number of measurements per participant is larger and 3) if the number of participants in trials is smaller.

\end{document}